\documentclass[prb,preprint,onecolumn,showpacs,preprintnumbers,amsmath,amssymb,superscriptaddress,floatfix]{revtex4-2}
\usepackage[T1]{fontenc}
\usepackage[ansinew]{inputenc}
\usepackage{graphics}
\usepackage{graphicx}
\usepackage{dcolumn}
\usepackage{bm}
\usepackage{amsthm}
\usepackage{amsmath}
\usepackage{amssymb}
\usepackage{diagbox}
\usepackage{hyperref}
\usepackage{url}
\usepackage{xcolor}

\begin{document}

\title{Signature of surface anisotropy in the spin-flip neutron scattering cross section of spherical nanoparticles: atomistic simulations and analytical theory}
 
\author{Michael P.\ Adams}\email[Electronic address: ]{michael.adams@uni.lu}
\affiliation{Department of Physics and Materials Science, University of Luxembourg, 162A~avenue de la Faiencerie, L-1511~Luxembourg, Grand Duchy of Luxembourg}

\author{Evelyn Pratami Sinaga}
\affiliation{Department of Physics and Materials Science, University of Luxembourg, 162A~avenue de la Faiencerie, L-1511~Luxembourg, Grand Duchy of Luxembourg}

\author{Hamid Kachkachi}
\affiliation{Universit\'{e} de Perpignan via Domitia, Laboratoire PROMES CNRS UPR8521, Rambla de la Thermodynamique, Tecnosud, F-66100~Perpignan, France}

\author{Andreas Michels}\email[Electronic address: ]{andreas.michels@uni.lu}
\affiliation{Department of Physics and Materials Science, University of Luxembourg, 162A~avenue de la Faiencerie, L-1511~Luxembourg, Grand Duchy of Luxembourg}


\begin{abstract}
We investigate the signature of magnetic surface anisotropy in nanoparticles in their spin-flip neutron scattering cross section. Taking into account the isotropic exchange interaction, an external magnetic field, a uniaxial or cubic magnetic anisotropy for the particle's core, and several models for the surface anisotropy (N\'{e}el, conventional, random), we compute the spin-flip small-angle neutron scattering (SANS) cross section from the equilibrium spin structures obtained using the Landau-Lifshitz equation. The sign of the surface anisotropy constant, which is related to the appearance of tangential- or radial-like spin textures, can be distinguished from the momentum-transfer dependence of the spin-flip signal. The data cannot be described by the well-known and often-used analytical expressions for uniformly magnetized spherical or core-shell particles, in particular at remanence or at the coercive field. Based on a second-order polynomial expansion for the magnetization vector field, we develop a novel minimal model for the azimuthally-averaged magnetic SANS cross section. The theoretical expression considers a general magnetization inhomogeneity and is not restricted to the presence of surface anisotropy. It is shown that the model describes very well our simulation data as well as more complex spin patterns such as vortex-like structures. Only seven expansion coefficients and some basis functions are sufficient to describe the scattering behavior of a very large number of atomic spins.
\end{abstract}

\date{\today}

\maketitle

 
\section{Introduction}

Research on magnetic nanoparticles is to a large extent driven by potential and existing applications in areas such as medicine, biology, and nanotechnology (see, e.g., Refs.~\cite{de2008applications,diebold2010applications,baetke2015applications,stark2015industrial,han2019applications,lakbenderdisch2021,BATLLE2022} and references therein). In many of the more application-oriented studies, the magnetization distribution (spin structure) of the nanoparticles is assumed to be uniform, i.e., the nanoparticles are considered to be in a single-domain state, where all the atomic magnetic moments are held in parallel by strong quantum-mechanical exchange forces. This approximation might be justified in some cases, e.g.\ for obtaining an initial overall understanding of a certain physical property or phenomenon, but there are also situations where it fails. A prominent example is magnetic hyperthermia on iron oxide nanoparticles~\cite{lakbenderdisch2021}, where the presence of microstructural defects---and the ensuing correlated spin disorder---gives rise to a strongly enhanced specific absorption rate as compared to the case of defect-free particles~\cite{lappas2019}. Therefore, understanding the spin structure of nanoparticles is not only of relevance from a fundamental science point of view but also from the standpoint of technological applications.

From the foregoing discussion one may realize that the steady development of both observational and computational methods to elucidate the magnetic microstructure of nanoparticles and to relate them to the macroscopic properties is an important task. In this respect we mention magnetic small-angle neutron scattering (SANS), which is probably the only experimental technique that is able to probe spin structures on the here-relevant mesoscopic length scale ($\sim$$1$$-$$1000 \, \mathrm{nm}$) and inside the volume of magnetic materials~\cite{michelsbook}. It is therefore not surprising that numerous experimental SANS investigations on nanoparticle systems including ferrofluids have been conducted to date~\cite{rmp2019,dirkreview2022}. An often reached conclusion is that the magnetization distribution inside the nanoparticles is not homogeneous but highly complex in the sense that a large variety of nonuniform, canted, vortex-type, or core-shell-type configurations are reported (e.g., \cite{suzuki2007,michels08epl,disch2012,kryckaprl2014,ijiri2014,guenther2014,maurer2014,dennis2015,feygenson2015,grutter2017,laura2017,oberdick2018,krycka2019,benderapl2019,bersweiler2019,zakutna2020,bender2020,suzuki2022,zakutna2023}). Numerical micromagnetic simulations play an increasingly important role in this context since they are able to predict the spin structures of nanoparticles and their related magnetic neutron scattering cross section and real-space correlation function. The magnetic ground state of a nanoparticle depends sensitively on many factors such as the particle size and shape, the presence of defects (e.g., vacancies, antiphase boundaries, surface anisotropy)~\cite{nedelkoski2017,lappas2019}, or simply on the magnetic interactions that are taken into account in the simulations. For instance, using Monte Carlo simulations of a discrete atomistic spin model, K\"ohler~et~al.~\cite{koehlerjac2021} have numerically studied the influence of antiphase boundaries in iron oxide nanoparticles on their spin structure. Instead of computing the magnetic SANS cross section, these authors used the Debye scattering equation to indirectly relate the internal spin disorder to the broadening of certain x-ray Bragg peaks. Vivas~et~al.~\cite{laura2020,evelynprb2023} carried out micromagnetic continuum calculations of the spin structure of iron nanoparticles and related a vortex-type magnetization configuration to certain signatures in the magnetic neutron scattering cross section and correlation function.

To further improve and advance the understanding of magnetic SANS, and its theoretical description using micromagnetic theory, it is important to realize the enormous complexity that is embodied in typical experimental SANS procedures. Quite commonly, experimental SANS data are presented as a plot of the azimuthally-averaged scattering intensity $I$ as a function of the magnitude of the scattering vector $q = |\mathbf{q}|$. The azimuthal averaging is frequently carried out along certain directions in $\mathbf{q}$~space, e.g., parallel or perpendicular to an externally applied magnetic field or over the full ($2\pi$) angular range on the two-dimensional (2D) detector. Although there is no principle difficulty in analyzing 2D SANS data (apart from an increased numerical effort), most of the time the 1D $I(q)$~data are analyzed, i.e., fitted to a certain scattering model or Fourier-transformed to obtain the correlation function. To finally arrive at the 1D $I(q)$~data involves a number of approximations and averaging processes that we will discuss in the following. 

A magnetic nanoparticle sample may be assumed to consist of a 3D distribution of nanoparticles that are rigidly embedded in some nonmagnetic matrix. The particles may be of different sizes and shapes, each particle is characterized by a certain magnetic anisotropy (e.g., of cubic and/or of uniaxial symmetry), even the type of microstructural defect or the surface anisotropy may vary between particles, and for dense assemblies magnetodipolar interactions between the particles are additionally important. For ferrofluids, which we do not consider here, additional interactions (e.g., of hydrodynamic origin) may become important. The experimental SANS cross section of such a system represents an average over all these features (particle-size and shape distributions, random easy-axis orientations of the particles, defects and surface anisotropy, dipolar interaction,...). Moreover, in small-angle approximation, the component of the scattering vector along the incident neutron beam is much smaller than the other two components, so that only correlations in the perpendicular plane are probed. Therefore, in the simulations, the computed three-dimensional SANS cross section needs to be projected onto the plane of the two-dimensional detector, which then yields the one-dimensional $I(q)$~curves by azimuthal averaging. In the course of all these steps, some information on the spin structure of the particles is lost, both in the simulations as well as in the experimental procedures.

Given the above-described enormous complexity involved in obtaining and analyzing experimental scattering data, one strategy to improve the current understanding is to systematically vary certain parameters in micromagnetic simulations and to track down their signatures in the randomly-averaged scattering signal. This approach is followed in the present work. We focus on the effect of strong surface anisotropy in nanoparticles and study the scattering signature of different functional dependencies for the magnetic behavior of the surface spins. Inspired by the obtained results and the related discussion, we derive a generally-applicable analytical formula for the magnetic SANS cross section.

More specifically, in this paper we employ atomistic simulations using the Landau-Lifshitz equation (LLE) to scrutinize the signature of surface anisotropy in magnetic nanoparticles in their spin-flip (sf) SANS cross section. The latter cross section can be routinely measured at many SANS beamlines using polarized neutrons; it is only composed of the Fourier components of the magnetization and possesses the advantage that the (unwanted) nuclear coherent SANS contribution---which can be quite large in nanoparticle systems---is absent. Therefore, the sf SANS cross section provides the most direct access to the spin microstructure of nanoparticles. In the simulations, we take into account the isotropic exchange interaction, an external magnetic field, a uniaxial or cubic magnetocrystalline anisotropy for the core of the nanoparticles, and most importantly different models for the anisotropy of the surface spins (N\'{e}el, conventional, random). As a central result of this study, we introduce a novel multi-nanoparticle power-series magnetization vector field model, which provides an analytical expression for the azimuthally-averaged magnetic SANS cross section. We emphasize that this analytical formula considers a general magnetization inhomogeneity, and is not necessarily restricted to surface anisotropy as the main mechanism to generate spin disorder. The theoretical expression is fitted to atomistic and coarse-grained micromagnetic simulation data.

The paper is organized as follows: In Sec.~\ref{sec2} we provide information on the atomistic simulations using the LLE [Sec.~\ref{sec2a}] and we display the expressions for the sf SANS cross section [Sec.~\ref{sec2b}]. The results of the numerical calculations are presented in Sec.~\ref{sec3a}, while Sec.~\ref{sec3b} provides a detailed discussion of the failure of the often-used structural core-shell-type form-factor models for the inhomogeneous spin structure of nanoparticles. In Section~\ref{sec4} and Appendix~\ref{appendixa} we introduce a multi-particle power-series analysis of the magnetic SANS cross section (involving all particles of an assembly). The theoretical model is used to interpret the present simulation data as well as scattering curves originating from more complex structures such as vortices in spheres. Section~\ref{sec5} summarizes the main findings of this study and provides an outlook on future challenges. Appendix~\ref{appendixb} shows results for the effect of the core-anisotropy symmetry (cubic versus uniaxial) on the sf SANS cross section. The Supplemental Material~\cite{michael2024sm} to this paper features several videos that display the SANS observables during the magnetization-reversal process for different sign combinations of the cubic/uniaxial core and surface anisotropy constants.

\section{Atomistic description of the nanoparticle spin structure and spin-flip SANS cross section}
\label{sec2}

\subsection{Atomistic spin Hamiltonian}
\label{sec2a}

\begin{figure}[tb!]
\centering
\resizebox{0.80\columnwidth}{!}{\includegraphics{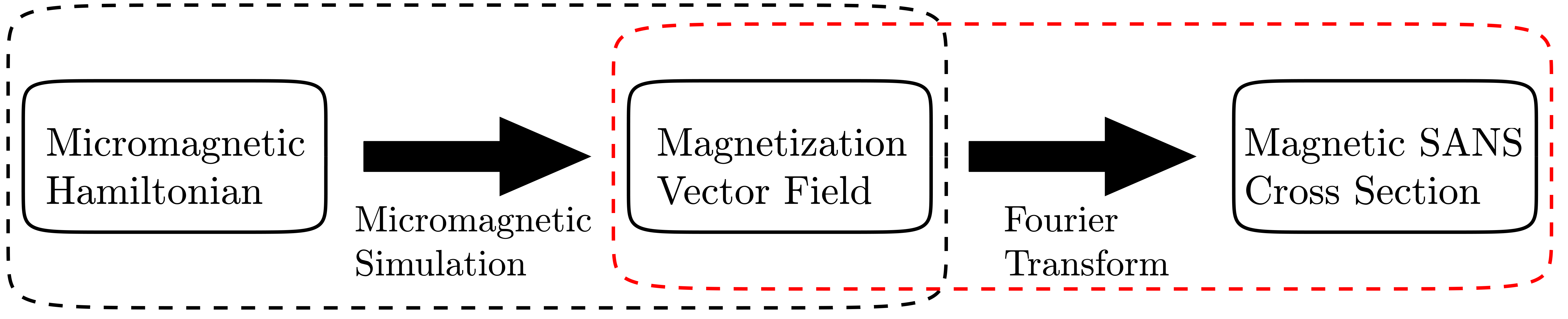}}
\caption{Basic concept of the connection between micromagnetic simulations and the magnetic SANS cross section.}
\label{fig1}
\end{figure}

Details of the atomistic simulation methodology of magnetic neutron scattering can be found in Ref.~\cite{adamsjacnum2022}. Here, we recall only the basic steps and relations in order to achieve a self-contained presentation. Figure~\ref{fig1} sketches the basic connection between atomistic or continuum micromagnetic simulations and the magnetic SANS cross section.

A spherical many-spin nanomagnet is viewed as a crystallite consisting of $\mathcal{N}$ atomic magnetic moments $\boldsymbol{\mu}_i = \mu_{\mathrm{a}} \mathbf{m}_i$, where $\mu_{\mathrm{a}}$ denotes the magnitude of the atomic magnetic moment and $\mathbf{m}_i$ is a unit vector specifying its orientation. We assume the spins to ``sit'' on a simple cubic lattice, so that $\mu_{\mathrm{a}} = M_{\mathrm{s}} a^3$, where $M_{\mathrm{s}}$ is the saturation magnetization of the material and $a$ the lattice constant. The spherical shape of the nanomagnet is cut from a simple cubic regular grid, and its radius $R$ is defined as $R = \frac{N-1}{2} a$, where the integer $N$ is the number of atoms on the side of the cubic grid. The magnetic state of the nanomagnet is investigated within an atomistic approach based on the following Hamiltonian~\cite{adamsjacnum2022}:
\begin{align}
\mathcal{H} &= \mathcal{H}_{{\mathrm{EX}}} + \mathcal{H}_{\mathrm{Z}} + \mathcal{H}_{\mathrm{A}} \nonumber \\
&= -\frac{1}{2} J \sum_{i,j \in{\mathrm{n.n.}}} \mathbf{m}_{i} \cdot \mathbf{m}_{j} - \mu_{a} \mathbf{B}_0 \cdot \sum_{i=1}^{\mathcal{N}} \mathbf{m}_{i} + \sum_{i=1}^{\mathcal{N}}\mathcal{H}_{\mathrm{A},i} ,
\label{eq:Ham-MSP}
\end{align}
where $\mathcal{H}_{{\mathrm{EX}}}$ is the nearest-neighbor (n.n.) exchange energy, with $J > 0$ the exchange parameter, $\mathcal{H}_{{\mathrm{Z}}}$ denotes the Zeeman energy with $\mathbf{B}_0$ the homogeneous externally applied magnetic field, and $\mathcal{H}_{{\mathrm{A}}}$ represents the magnetic anisotropy energy. For the core spins, we assume the anisotropy to be either of uniaxial (``u'') or cubic (``c'') symmetry, while for the surface spins we adopt several models. More specifically, $\mathcal{H}_{\mathrm{A},i}$ is expressed as follows:
\begin{equation}
\mathcal{H}_{\mathrm{A},i} = \begin{cases}
-K_{\mathrm{u}} \left(\mathbf{m}_{i} \cdot \mathbf{e}_{\mathrm{A}} \right)^{2} \hspace{0.5cm} \mathrm{or} \hspace{0.5cm} K_{\mathrm{c}} \left( m_{i,x}^2 m_{i,y}^2 + m_{i,x}^2 m_{i,z}^2 + m_{i,y}^2 m_{i,z}^2 \right) , & i \in{\mathrm{core}}
\\ \\ \frac{1}{2} K_{\mathrm{s}} {\displaystyle \sum_{j\in{\mathrm{n.n.}}}} \left(\mathbf{m}_{i}\cdot\mathbf{u}_{ij}\right)^{2} \hspace{0.5cm} \mathrm{or} \hspace{0.5cm} - \frac{1}{2} K_{\mathrm{s}} \left( \mathbf{m}_{i} \cdot \mathbf{n} \right)^{2} , & i\in\mathrm{surface} ,
\end{cases}
\label{eq:HamUA-NSA}
\end{equation}
where $K_{\mathrm{u}}$, $K_{\mathrm{c}}$, and $K_{\mathrm{s}}$ denote, respectively, the uniaxial or cubic core and surface anisotropy constants, $\mathbf{e}_{\mathrm{A}}$ is a unit vector along the easy axis of the core, $\mathbf{u}_{ij} = (\mathbf{r}_i - \mathbf{r}_j) / \|\mathbf{r}_i - \mathbf{r}_j \|$ is a unit vector connecting the nearest-neighbor spins $i$ and $j$, and $\textbf{n}$ is the unit normal vector on the surface of the spherical particle. The surface spins are defined as those spins which have a coordination number less than six (simple cubic lattice). The particular model with the $\mathbf{u}_{ij}$~vectors is the one proposed by N\'{e}el~\cite{nee54jpr}, while the expression involving the unit normal vector $\mathbf{n}$ on the surface is an often-used phenomenological way to describe surface anisotropy effects. Additionally, we consider the case of a random surface anisotropy, where we take the $\mathbf{u}_{ij}$ as random vectors. In the following, these three models will be denoted as, respectively, the N\'{e}el model (NM), the conventional model (CM), and the random surface anisotropy model (RM). The anisotropy constants in Eq.~(\ref{eq:HamUA-NSA}) can assume positive as well as negative signs. The magnetodipolar interaction has been ignored in our simulations. This is motivated by the numerical complexity of this energy term, in particular for atomistic simulations (here, for a $8 \, \mathrm{nm}$~diameter particle the number of spins is $\mathcal{N} = 11363$), and by the expectation that it is of minor relevance for smaller-sized nanomagnets~\cite{koehlerjac2021,hertel2021}.

The dynamics of each individual magnetic moment $\mathbf{m}_{i}$ is described by the Landau-Lifshitz equation (LLE)~\cite{berkovinkronparkinhandbook07}:
\begin{equation}
\frac{d{\mathbf{m}}_i}{dt} = - \gamma \, \mathbf{m}_{i} \times \mathbf{B}_{i}^{\mathrm{eff}} - \alpha \, \mathbf{m}_{i} \times (\mathbf{m}_{i} \times \mathbf{B}_{i}^{\mathrm{eff}}) ,
\label{eq:LLLE}
\end{equation}
where $\gamma$ is the gyromagnetic ratio and $\alpha$ denotes the damping constant. The effective magnetic field $\mathbf{B}_{i}^{\mathrm{eff}}$ acting on spin ``$i$'' is obtained as the functional derivative of $\mathcal{H}$ [Eq.~(\ref{eq:Ham-MSP})] with respect to $\mathbf{m}_{i}$, i.e., $\mathbf{B}_{i}^{\mathrm{eff}} = - \mu_{\mathrm{a}}^{-1} \delta \mathcal{H} / \delta \mathbf{m}_{i}$. The LLE is then solved numerically by using the explicit Euler forward-projection method~\cite{bavnas2005numerical}. At each value of the external field and for given materials parameters, atomistic simulations of the spin structure and of the ensuing magnetic neutron scattering cross section were carried out for typically $256$ random orientations of the core-anisotropy axes of the particle with respect to the field $\mathbf{B}_0$, which defines the $z$~direction of a Cartesian laboratory coordinate frame. More specifically, once the lattice orientation has been randomly selected, the easy-axis orientation of the particle's core and the distribution of the surface anisotropy are fixed. The whole system (core plus surface anisotropy) is then randomly rotated relative to $\mathbf{B}_0$. The simulations were carried out by starting from a large positive (saturating) field of about $10 \, \mathrm{T}$ and then the field was reduced in steps of typically $30 \, \mathrm{mT}$.

In our simulations we used the following parameters:~atomic magnetic moment of bcc iron $\mu_{\mathrm{a}} = 2.22 \, \mu_\mathrm{B}$ (with $\mu_\mathrm{B}$ the Bohr magneton), lattice constant $a = 0.287 \, \mathrm{nm}$, exchange constant $J = 3.01 \times 10^{-22} \, \mathrm{J/atom}$, damping constant $\alpha = 3 \times 10^{11} \, \mathrm{T}^{-1}\mathrm{s}^{-1}$, gyromagnetic constant $\gamma = 1.76 \times 10^{11} \, \mathrm{T}^{-1}\mathrm{s}^{-1}$, and an integration time step of $h_{\mathrm{t}} = 5 \, \mathrm{fs}$. For the uniaxial ($K_{\mathrm{u}}$) and cubic ($K_{\mathrm{c}}$) core anisotropy constants, we used the value of $K_{\mathrm{u}} = K_{\mathrm{c}} = + 5.67 \times 10^{-25} \, \mathrm{J/atom}$. Experimental $K_{\mathrm{s}}$~values for nanoparticles and thin films can be found in Refs.~\cite{gradmann86,BATLLE2022,ohandley}. A value of $K_{\mathrm{s}} = 5.22 \times 10^{-21} \, \mathrm{J/atom}$---which we use in the simulations---has been estimated in Ref.~\cite{kacdim02prb,kacbon06prb} for a $4 \, \mathrm{nm}$-sized fcc Cobalt particle. Note, however, that we vary the sign of $K_{\mathrm{s}}$ [compare to Eq.~(\ref{eq:HamUA-NSA})].

For the calculation of the sf SANS cross section $d\Sigma_{\mathrm{sf}} / d\Omega$ [see Eq.~(\ref{eq:equation1}) below], it is necessary to compute the discrete Fourier transform of all ${\mathbf{m}}_i$ belonging to the spherical nanomagnet. Using $\boldsymbol{\mu}_{i} = \mu_{a} \mathbf{m}_{i}$, the discrete-space Fourier transform is computed as: 
\begin{align}
\widetilde{\mathbf{M}}(\mathbf{q}) = \frac{\mu_a}{(2\pi)^{3/2}} \sum_{i=1}^{\mathcal{N}} \mathbf{m}_i \exp\left(-\mathrm{i} \mathbf{q} \cdot \mathbf{r}_i \right) ,
\label{discreteFT}
\end{align}
where $\mathbf{r}_i$ is the location point of the $i$th spin and $\mathbf{q}$ represents the wave vector (scattering vector). Equation~(\ref{discreteFT}) establishes the relation between the outcome of the spin-configuration simulations, namely the $\mathbf{m}_i$, and the sf SANS cross section $d\Sigma_{\mathrm{sf}} / d\Omega$.

\subsection{Spin-flip SANS cross section}
\label{sec2b}

\begin{figure}[tb!]
\centering
\resizebox{0.80\columnwidth}{!}{\includegraphics{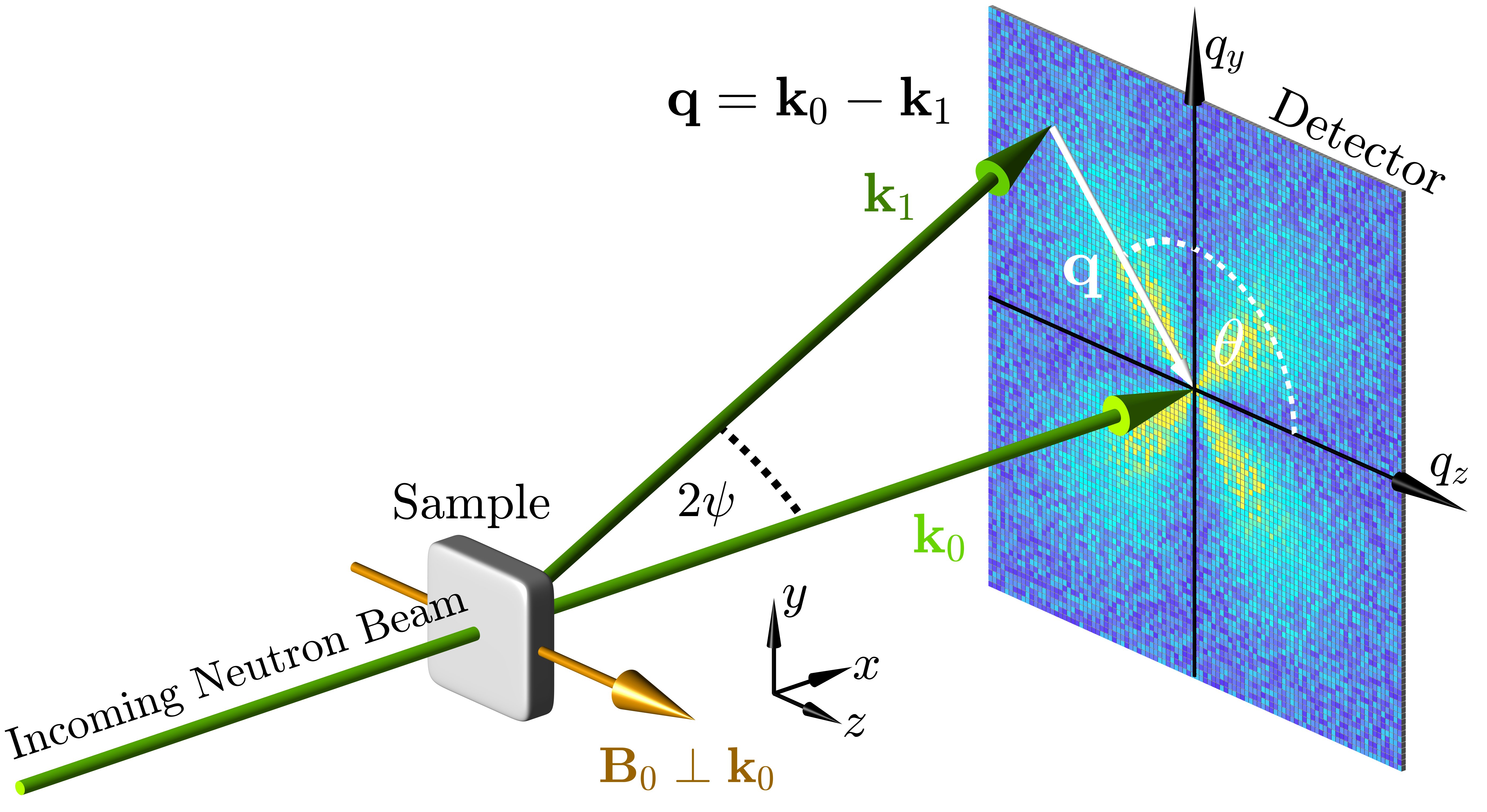}}
\caption{Sketch of the neutron scattering geometry. The neutron optical elements (polarizer, spin flipper, analyzer) that are required to measure the spin-flip SANS cross section are not drawn. The applied magnetic field $\mathbf{B}_0 \parallel \mathbf{e}_z$ is perpendicular to the wave vector $\mathbf{k}_0 \parallel \mathbf{e}_x$ of the incident neutron beam ($\mathbf{B}_0 \perp \mathbf{k}_0$). The momentum-transfer or scattering vector $\mathbf{q}$ is defined as the difference between $\mathbf{k}_0$ and $\mathbf{k}_1$, i.e., $\mathbf{q} = \mathbf{k}_0 - \mathbf{k}_1$. SANS is usually implemented as elastic scattering ($k_0 = k_1 = 2\pi / \lambda$), and the component of $\mathbf{q}$ along the incident neutron beam, here $q_x$, is much smaller than the other two components, so that $\mathbf{q} \cong \{ 0, q_y, q_z \} = q \{ 0, \sin\theta, \cos\theta \}$. This demonstrates that SANS probes predominantly correlations in the plane perpendicular to the incident beam. For elastic scattering, the magnitude of $\mathbf{q}$ is given by $q = (4\pi / \lambda) \sin(\psi)$, where $\lambda$ denotes the mean wavelength of the neutrons and $2\psi$ is the scattering angle. The angle $\theta = \angle(\mathbf{q}, \mathbf{B}_0)$ is used to describe the angular anisotropy of the recorded scattering pattern on the two-dimensional position-sensitive detector.}
\label{fig2}
\end{figure}

The quantity of interest in experimental SANS studies is the elastic magnetic differential scattering cross section, which is usually recorded on a two-dimensional position-sensitive detector. Progress in SANS instrumentation allows one to routinely measure the spin-flip (sf) SANS cross section $d \Sigma_{\mathrm{sf}} / d \Omega$, which does not contain the nuclear coherent scattering signal. For the most commonly used scattering geometry in magnetic SANS experiments, where the applied magnetic field $\mathbf{B}_0 \parallel \mathbf{e}_z$ is perpendicular to the wave vector $\mathbf{k}_0 \parallel \mathbf{e}_x$ of the incident neutrons (see Fig.~\ref{fig2}), $d \Sigma_{\mathrm{sf}} / d \Omega$ can be written as~\cite{michelsbook}:
\begin{eqnarray}
\frac{d \Sigma_{\mathrm{sf}}}{d \Omega}(\mathbf{q}) = \frac{8 \pi^3}{V} b_{\mathrm{H}}^2 \left( |\widetilde{M}_x|^2 + |\widetilde{M}_y|^2 \cos^4\theta \right. \nonumber \\ \left. + |\widetilde{M}_z|^2 \sin^2\theta \cos^2\theta - (\widetilde{M}_y \widetilde{M}_z^{\ast} + \widetilde{M}_y^{\ast} \widetilde{M}_z) \sin\theta \cos^3\theta \right) ,
\label{eq:equation1}
\end{eqnarray}
where $V$ is the scattering volume, $b_{\mathrm{H}} = 2.91 \times 10^8 \, \mathrm{A}^{-1}\mathrm{m}^{-1}$ is the magnetic scattering length in the small-angle regime (the atomic magnetic form factor is approximated by $1$ since we are dealing with forward scattering), $\widetilde{\mathbf{M}}(\mathbf{q}) = \{ \widetilde{M}_x(\mathbf{q}), \widetilde{M}_y(\mathbf{q}), \widetilde{M}_z(\mathbf{q}) \}$ represents the Fourier transform of the magnetization vector field $\mathbf{M}(\mathbf{r}) = \{ M_x(\mathbf{r}), M_y(\mathbf{r}), M_z(\mathbf{r}) \}$, $\theta$ denotes the angle between $\mathbf{q}$ and $\mathbf{B}_0$, and the asterisk ``$*$'' stands for the complex conjugate. Note that in the perpendicular scattering geometry the Fourier components are evaluated in the plane $q_x = 0$ (compare to Fig.~\ref{fig2}). In writing down Eq.~(\ref{eq:equation1}) we have ignored the polarization-dependent chiral scattering term. This contribution has also been calculated in our numerical procedure and, as expected, its magnitude is found to be typically $2$$-$$3$ orders of magnitude smaller than the other terms in Eq.~(\ref{eq:equation1}).

The numerically computed sf SANS cross sections that are shown in this paper correspond to the following average:
\begin{eqnarray}
\frac{d \Sigma_{\mathrm{sf}}}{d \Omega}(\mathbf{q}) = \left\langle \frac{d \Sigma_{\mathrm{sf}}}{d \Omega}(\mathbf{q}) \right\rangle_{\mathrm{EA}} = \frac{1}{\mathcal{K}} \sum_{k=1}^{\mathcal{K}} \frac{d \Sigma_{{\mathrm{sf}},k}}{d \Omega}(\mathbf{q}),
\label{eq:equation1average}
\end{eqnarray}
where $d \Sigma_{{\mathrm{sf}},k} / d \Omega$ represents (for fixed $K_{\mathrm{u}}$, $K_{\mathrm{c}}$, $K_{\mathrm{s}}$, $B_0$) the sf SANS cross section for a particular easy-axis (``EA'') orientation of the particle core (referred to index ``$k$''), and $\mathcal{K} = 256$ denotes the number of random configurations (total number of particles). Equation~(\ref{eq:equation1average}) implies the absence of interparticle interactions.

It is often convenient to average the two-dimensional SANS cross section $\frac{d \Sigma_{\mathrm{sf}}}{d \Omega}(\mathbf{q}) = \frac{d \Sigma_{\mathrm{sf}}}{d \Omega}(q_y, q_z) = \frac{d \Sigma_{\mathrm{sf}}}{d \Omega}(q, \theta)$ along certain directions in $\mathbf{q}$-space, e.g.\ parallel ($\theta = 0$) or perpendicular ($\theta = \pi/2$) to the applied magnetic field, or even over the full angular $\theta$~range. In the following, we consider the $2\pi$~azimuthally-averaged sf SANS cross section
\begin{eqnarray}
\label{aziaverage}
I_{\mathrm{sf}}(q) = \frac{1}{2\pi} \int_0^{2\pi} \frac{d \Sigma_{\mathrm{sf}}}{d \Omega}(q,\theta) \, d\theta .
\end{eqnarray}
The influence of a distribution of particle sizes on the magnetic SANS observables has been modeled using a lognormal probability distribution function, which is defined as~\cite{krill98}:
\begin{align}
w(D) = \frac{1}{\sqrt{2\pi D^2  
\operatorname{ln}\left(1 + \frac{\sigma^2}{\mu^2}\right)} } \exp\left(- 
\frac{\mathrm{ln}^2\left(\frac{D}{\mu}\sqrt{1 + \frac{\sigma^2}{\mu^2}}\right)}{2\operatorname{ln}\left(1 + \frac{\sigma^2}{\mu^2}\right)}
\right),
\end{align}
where $\mu$ denotes the expectation value and $\sigma^2$ is the variance, such that:
\begin{align}
\mu &= \int_{0}^{\infty} w(D) D d D > 0 ,
\\
\sigma^2 &= \int_{0}^{\infty} w(D) (D - \mu)^2 d D .
\end{align}
The corresponding median $\mu^{\ast}$ is determined by the following relation:
\begin{align}
\int_{0}^{\mu^{\ast}} w(D) d D = \frac{1}{2} \hspace{0.5cm} \mathrm{and} \hspace{0.5cm} \mu^{\ast} = \frac{\mu^2}{\sqrt{\mu^2 + \sigma^2}} .
\end{align}
For given values of $\mu$ and $\sigma$, the averaged sf SANS cross section $\langle ... \rangle$ is computed as:
\begin{align}
\left \langle \frac{d\Sigma_{\mathrm{sf}}}{d\Omega} \right \rangle &  = \frac{1}{V} \sum_{\ell = 1}^{L} \frac{d\Sigma_{{\mathrm{sf}},\ell}}{d\Omega} P_\ell V_{\ell} ,
\label{lognorfunc}
\\
V_{\ell} &=  \frac{4\pi R_{\ell}^3}{3} ,
\\
V &=  \sum_{\ell=1}^{L} P_{\ell} V_{\ell} ,
\end{align}
where $P_\ell$ denotes the probability (weight) related to the particle-size class $D_\ell = 2 R_{\ell}$ (diameter) by
\begin{align}
P_{\ell} = \int_{D_{\ell}-\Delta D/2}^{D_{\ell}+\Delta D/2} w(D) d D .
\end{align}
The $d \Sigma_{{\mathrm{sf}},\ell} / d\Omega$ in Eq.~(\ref{lognorfunc}) correspond to the randomly-averaged sf cross section of the size class $\ell$, computed according to Eq.~(\ref{eq:equation1average}).

\section{Results and Discussion}
\label{sec3}

\subsection{Atomistic simulations: spin structures and spin-flip SANS cross section}
\label{sec3a}

\begin{figure}[tb!]
\centering
\resizebox{1.0\columnwidth}{!}{\includegraphics{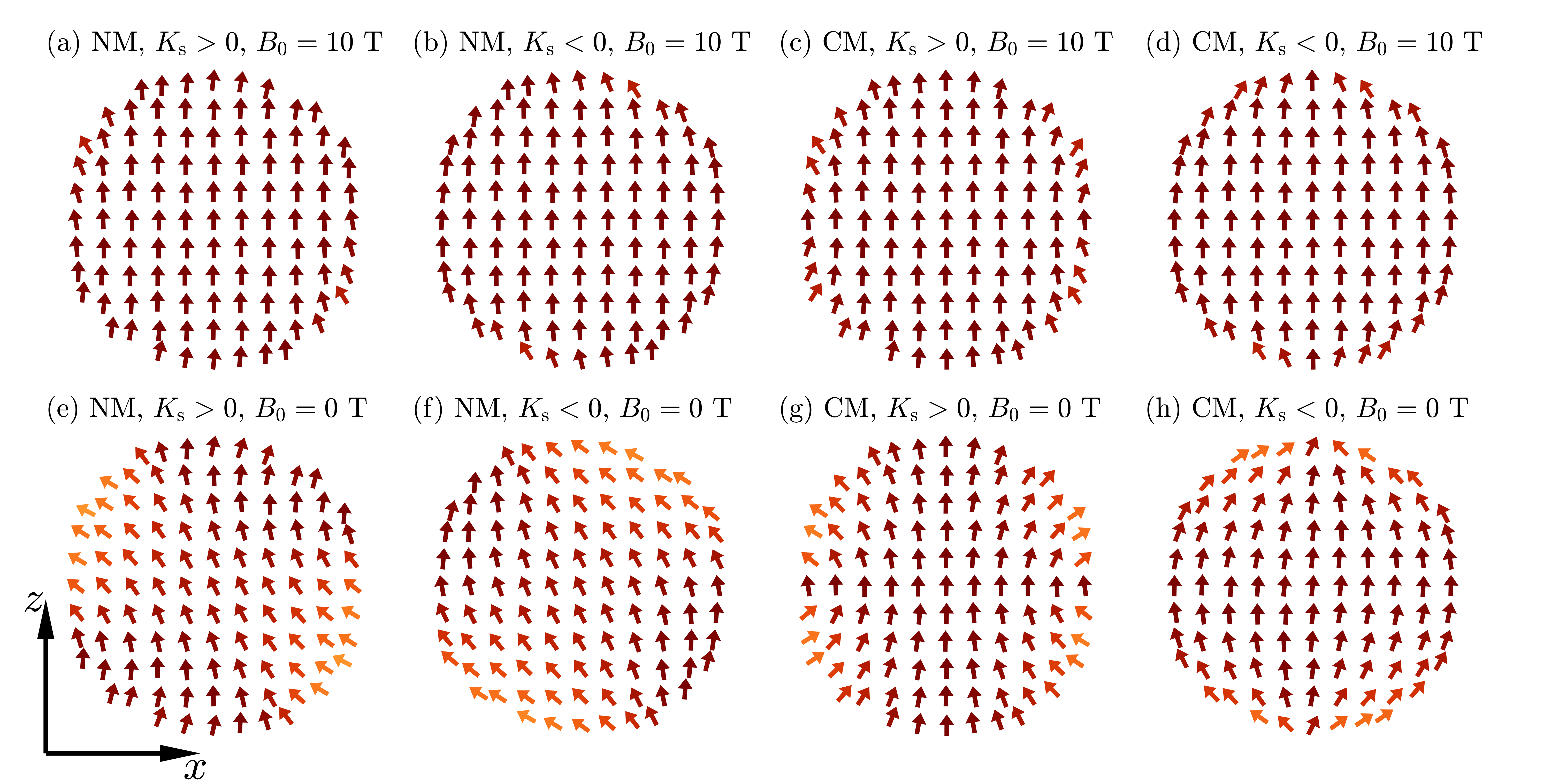}}
\caption{Spin structures (snapshots) of $8 \, \mathrm{nm}$ diameter nanoparticles at selected applied magnetic fields and for different sign combinations of the cubic core ($K_{\mathrm{c}}$) and surface ($K_{\mathrm{s}}$) anisotropy constants. The images represent projections of the three-dimensional spin structure into the plane $y=0$. $K_{\mathrm{c}} = + 5.67 \times 10^{-25} \, \mathrm{J/atom}$ in all simulations and $|K_{\mathrm{s}}| = 5.22 \times 10^{-21} \, \mathrm{J/atom}$, with the sign of $K_{\mathrm{s}}$ changing (see insets). NM~$=$~N\'{e}el model, CM~$=$~conventional model, RM~$=$~random surface anisotropy model. (a)$-$(d) near saturation ($B_0 = 10 \, \mathrm{T}$); (e)$-$(h) at remanence ($B_0 = 0 \, \mathrm{T}$). All spin structures were previously saturated along the $z$~direction.}
\label{fig3}
\end{figure}

Figure~\ref{fig3} shows examples for numerically-computed spin structures at remanence and saturation and for different sign combinations of the cubic core ($K_{\mathrm{c}}$) and surface ($K_{\mathrm{s}}$) anisotropy constants. The magnitudes of $K_{\mathrm{c}}$ and $K_{\mathrm{s}}$ are constant in these simulations, only the sign of $K_{\mathrm{s}}$ changes. Near saturation ($B_0 = 10 \, \mathrm{T}$, upper row in Fig.~\ref{fig3}), the particles are essentially uniformly magnetized, although some ``wiggling'' of the spins near the surface occurs due to the strong surface anisotropy. At lower fields (remanence, lower row in Fig.~\ref{fig3}), larger spin inhomogeneities appear throughout the volume of the particle. It is seen that the spins near the surface of the nanoparticle have a more radial orientation when $K_{\mathrm{s}} > 0$, while they are more tangentially oriented when $K_{\mathrm{s}} < 0$. This can be understood by inspecting the expression for the surface anisotropy energy [Eq.~(\ref{eq:HamUA-NSA})]; for $K_{\mathrm{s}} > 0$, the $\mathbf{m}_{i} \perp \mathbf{u}_{ij}$ and $\mathbf{m}_{i} \parallel \mathbf{n}$ orientations are energetically preferred, whereas $\mathbf{m}_{i} \parallel \mathbf{u}_{ij}$ and $\mathbf{m}_{i} \perp \mathbf{n}$ for $K_{\mathrm{s}} < 0$.

\begin{figure}[tb!]
\centering
\resizebox{1.0\columnwidth}{!}{\includegraphics{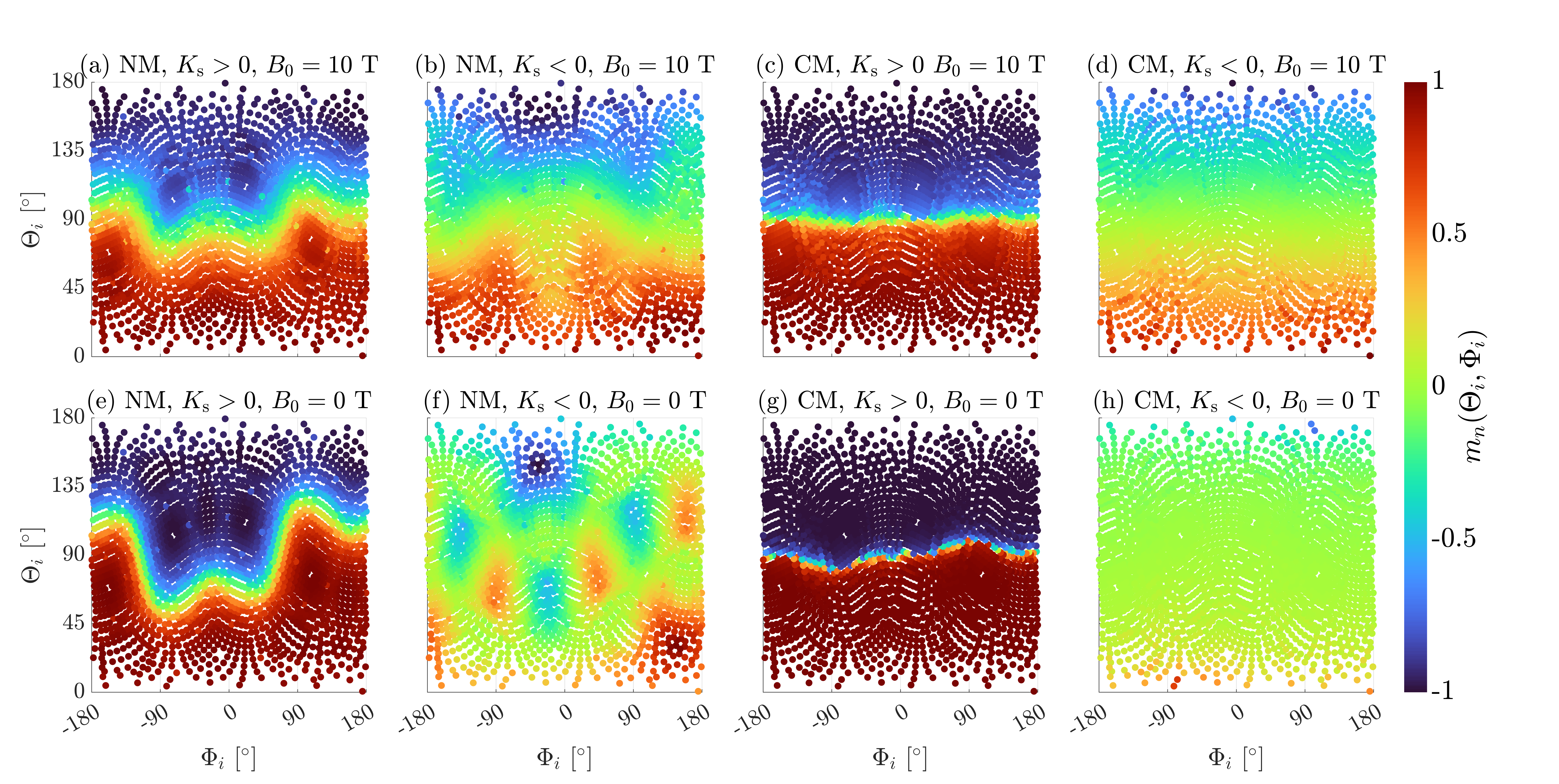}}
\caption{Alternative representation of the spin structure data shown in Fig.~\ref{fig3}. Plotted is the normal component of the spins at the surface, $m_n(\Theta_i, \Phi_i)$ [Eq.~(\ref{eq:SurfaceNormalMagnetizationComponent})], for the N\'{e}el (NM) and conventional (CM) model at $B_0 = 10 \, \mathrm{T}$ and at $B_0 = 0 \, \mathrm{T}$ both for $K_{\mathrm{s}} > 0$ and $K_{\mathrm{s}} < 0$. $\Theta_i$ and $\Phi_i$ denote, respectively, the polar and azimuthal angles. Same parameters and notation as in Fig.~\ref{fig3} (see insets).}
\label{fig4}
\end{figure}

Figure~\ref{fig4} displays a spherical map of the normal component of the spins at the surface, $m_n$, corresponding to the spin structures shown in Fig.~\ref{fig3}. We compute $m_n$ as:
\begin{align}
m_n(\Theta_i, \Phi_i) = \mathbf{m}_i \cdot \frac{\mathbf{r}_i}{\|\mathbf{r}_i\|} ,
\label{eq:SurfaceNormalMagnetizationComponent}
\end{align}
where the position vector $\mathbf{r}_i$ is expressed using the spherical coordinates $\Theta_i,\Phi_i$ as:
\begin{align}
\mathbf{r}_i = r_i \begin{Bmatrix}
\sin\Theta_i \cos\Phi_i , &
\sin\Theta_i \sin\Phi_i , &
\cos\Theta_i
\end{Bmatrix}
 \quad \mathrm{for} \quad 0.94 R \le r_i \le R .
\end{align}
$m_n(\Theta_i, \Phi_i)$ may be interpreted as a fictitious magnetic surface charge density on the spherical boundary surface in the direction that is specified by the polar angle $\Theta_i$ and the azimuthal angle $\Phi_i$. The images in Fig.~\ref{fig4} highlight, as described in the following, the effects of changing the sign of $K_{\mathrm{s}}$ on the spin textures at the surface of the nanoparticle. For more radial-like surface spins, corresponding to $K_{\mathrm{s}} > 0$, and for both the NM and CM model [Fig.~\ref{fig4}(a), (c), (e), (g)], we observe a relatively sharp separation of positive and negative $m_n(\Theta_i, \Phi_i)$~values around the equatorial line ($\Theta_i = 90^{\circ}$). Roughly speaking, the spins point radially outward in the upper hemisphere ($0^{\circ} \leq \Theta_i \leq 90^{\circ}$), while they point radially inward in the lower hemisphere ($90^{\circ} \leq \Theta_i \leq 180^{\circ}$). For the case $K_{\mathrm{s}} < 0$ [Fig.~\ref{fig4}(b), (d), (f), (h)], we find more extended regions on this map where $m_n(\Theta_i, \Phi_i)$ is close to zero (corresponding to tangential textures). For instance, for the CM model at zero field [Fig.~\ref{fig4}(h)], there is nearly no radial spin component; increasing the field to $10 \, \mathrm{T}$ [Fig.~\ref{fig4}(d)] gives rise to radial components, since $\mathbf{B}_0$ is applied parallel to the $z$~direction ($\Theta_i = 0^{\circ}, \Phi_i = 0^{\circ}$). For the NM model and $K_{\mathrm{s}} < 0$, we obtain a more complex behavior: $m_n(\Theta_i, \Phi_i)$ at zero field [Fig.~\ref{fig4}(f)] exhibits an ``oscillatory'' pattern of tangential and radial surface spin regions, which changes with increasing field [Fig.~\ref{fig4}(b)] to the expected structure that is dominated by radially oriented spins.

\begin{figure}[tb!]
\centering
\resizebox{1.0\columnwidth}{!}{\includegraphics{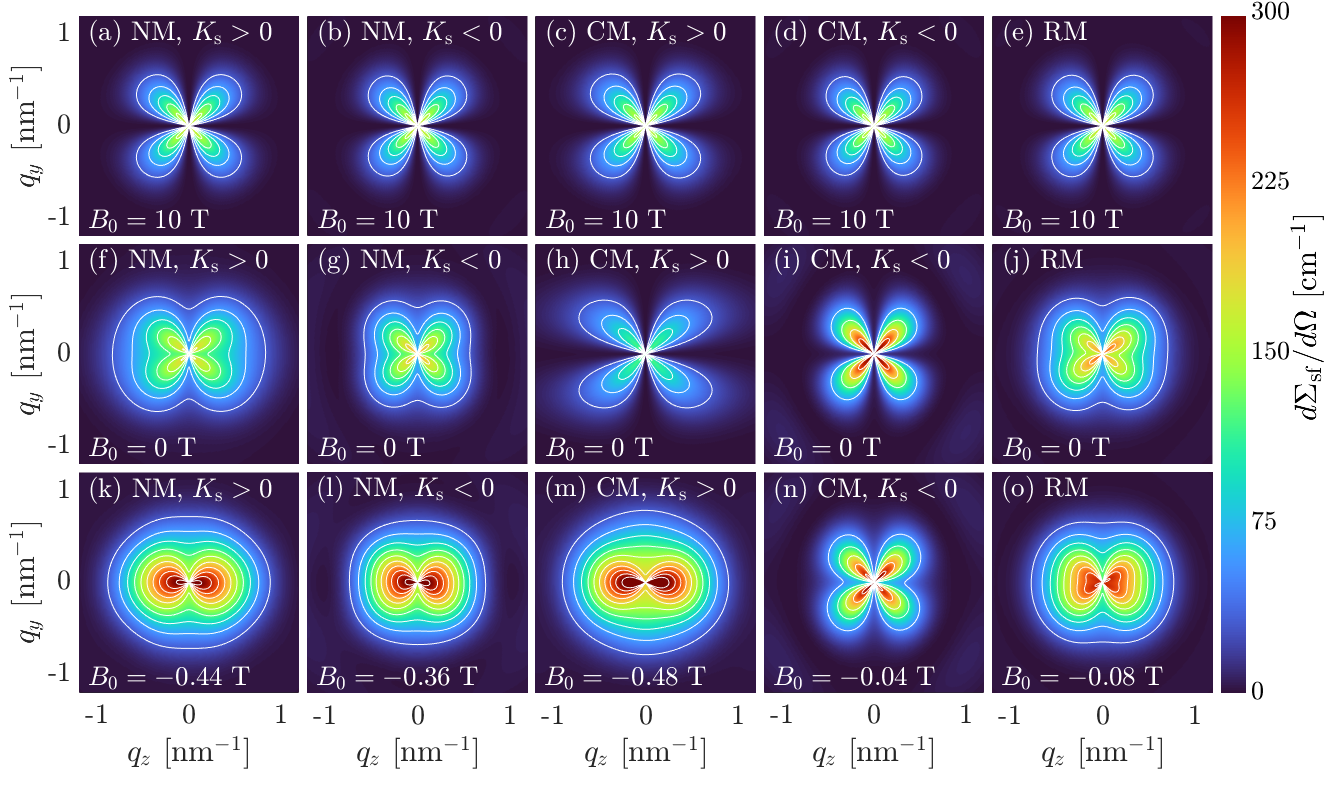}}
\caption{Numerically-computed two-dimensional spin-flip SANS cross sections $d \Sigma_{{\mathrm{sf}}} / d\Omega$ [Eq.~(\ref{eq:equation1})] at selected applied magnetic fields and for different sign combinations of the cubic core ($K_{\mathrm{c}}$) and surface ($K_{\mathrm{s}}$) anisotropy constants (linear color scale, $\mathbf{B}_0 \parallel \mathbf{e}_z \perp \mathbf{k}_0$). $K_{\mathrm{c}} = + 5.67 \times 10^{-25} \, \mathrm{J/atom}$ in all simulations and $|K_{\mathrm{s}}| = 5.22 \times 10^{-21} \, \mathrm{J/atom}$, with the sign of $K_{\mathrm{s}}$ changing (same notation as in Fig.~\ref{fig3}, see insets). (a)$-$(e) near saturation ($B_0 = 10 \, \mathrm{T}$); (f)$-$(j) at remanence ($B_0 = 0 \, \mathrm{T}$); (k)$-$(o) at the respective coercive field. The particle diameter is $D = 2R = 8 \, \mathrm{nm}$.}
\label{fig5}
\end{figure}

\begin{figure}[tb!]
\centering
\resizebox{1.0\columnwidth}{!}{\includegraphics{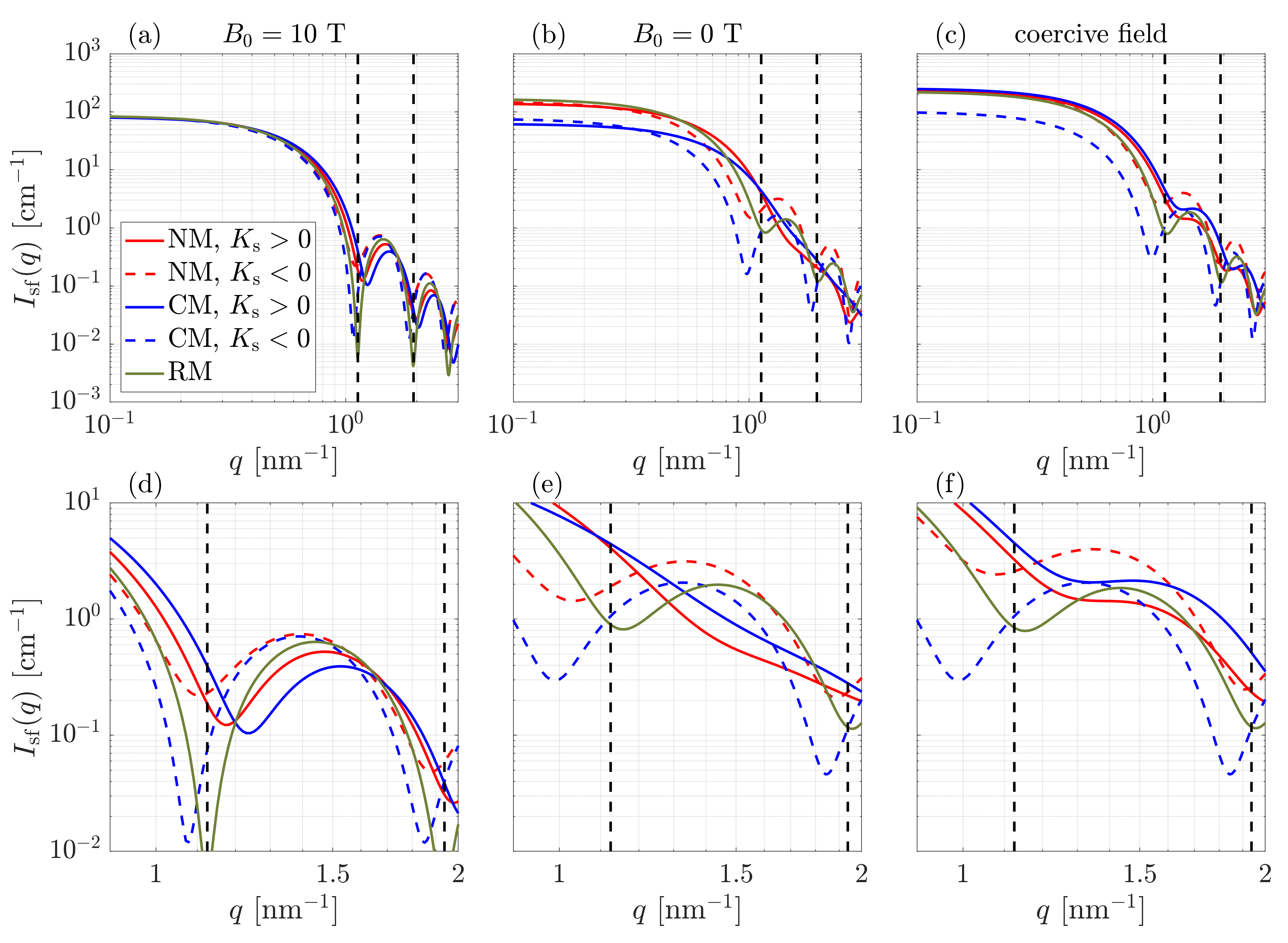}}
\caption{$2\pi$~azimuthally-averaged spin-flip SANS cross sections $I_{\mathrm{sf}}(q)$ (data from Fig.~\ref{fig5}) (log-log scales). (a)~near saturation ($B_0 = 10 \, \mathrm{T}$); (b)~at remanence ($B_0 = 0 \, \mathrm{T}$); (c)~at the respective coercive field. (d)$-$(f) display an enlarged region of the respective plots in (a)$-$(c). The vertical dashed lines mark the first two minima of the form factor of a homogeneous sphere ($I_{\mathrm{sf}}(q) \sim [j_1(qR)/(qR)]^2$ with a first zero at $q_1^{\mathrm{sat}} \cong 9/D = 1.125 \;\mathrm{nm}^{-1}$ and a second zero at $q_2^{\mathrm{sat}} \cong 15.5/D = 1.9375 \;\mathrm{nm}^{-1}$). The particle diameter is $D = 2R = 8 \, \mathrm{nm}$.}
\label{fig6}
\end{figure}

\begin{figure}[tb!]
\centering
\resizebox{0.80\columnwidth}{!}{\includegraphics{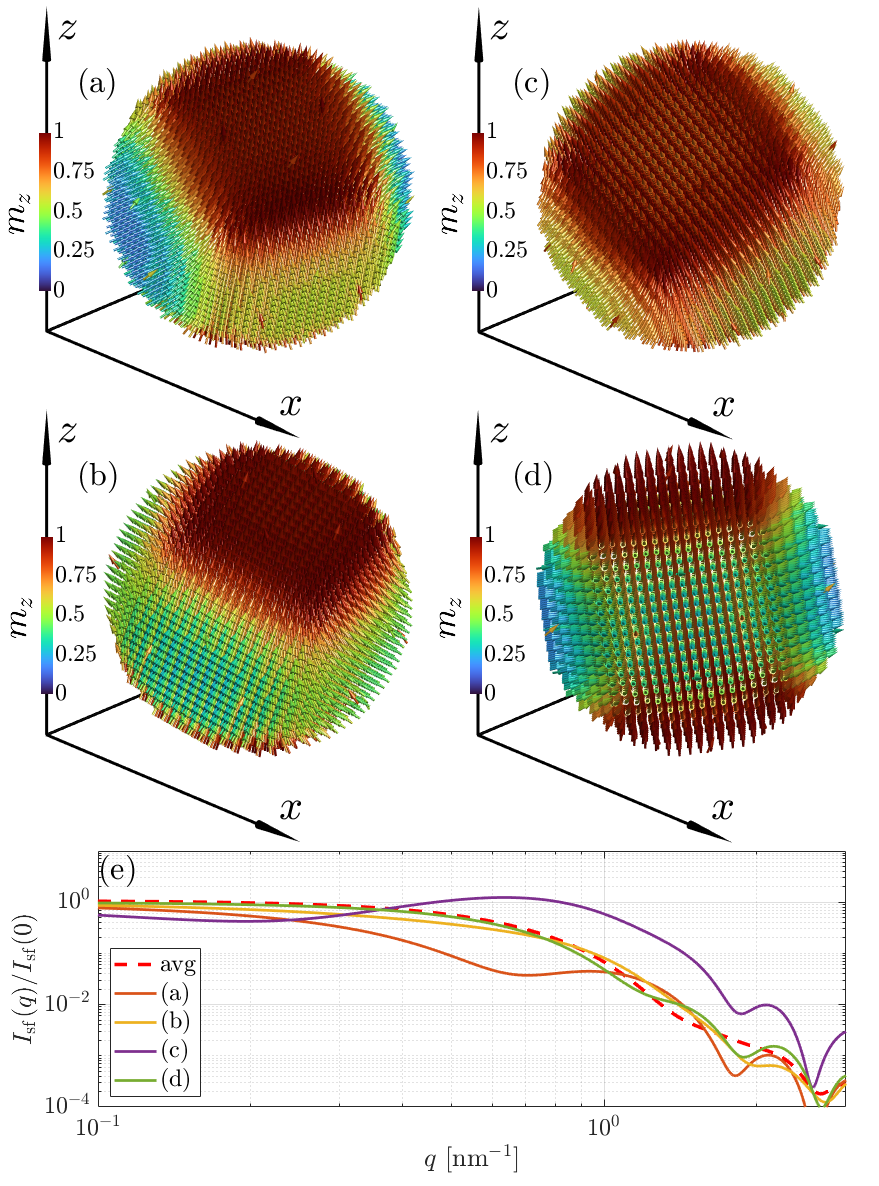}}
\caption{(a)$-$(d)~Selected real-space spin structures at $B_0 = 0 \, \mathrm{T}$ for the NM model with $K_{\mathrm{s}} > 0$ ($D = 2R = 8 \, \mathrm{nm}$). These structures differ in the their orientations of the core-anisotropy axes relative to $\mathbf{B}_0 \parallel \mathbf{e}_z$. (e)~Normalized $2\pi$~azimuthally-averaged spin-flip SANS cross sections $I_{\mathrm{sf}}(q)$ of the four spin structures shown in (a)$-$(d) along with the $I_{\mathrm{sf}}(q)$ (avg) of an ensemble of $256$ randomly-oriented particles (see inset) (log-log scale) [compare the definition of the sf SANS cross section, Eq.~(\ref{eq:equation1average})]. See also the corresponding movie in the Supplemental Material~\cite{michael2024sm}.}
\label{fig7}
\end{figure}

\begin{figure}[tb!]
\centering
\resizebox{1.0\columnwidth}{!}{\includegraphics{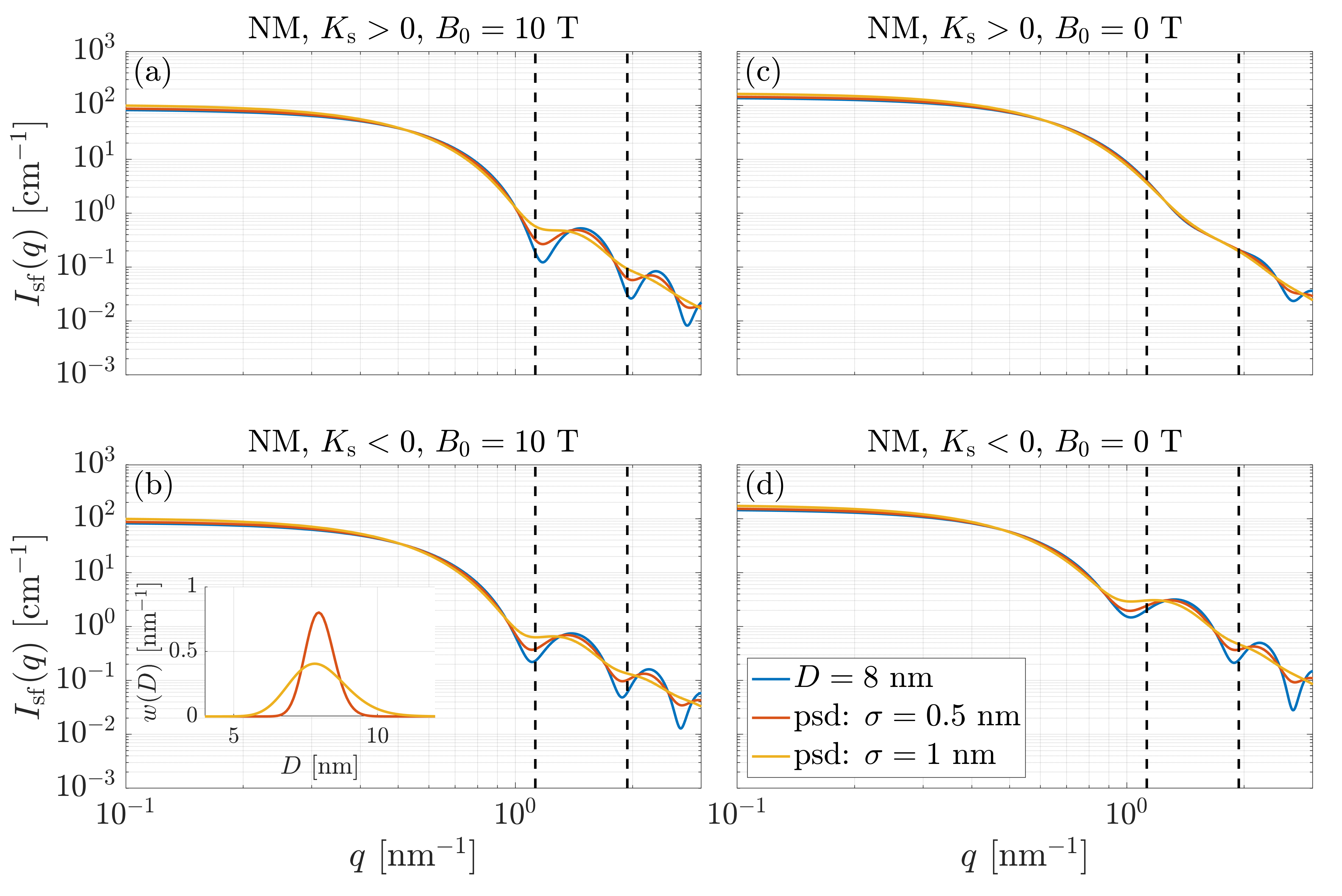}}
\caption{Effect of a particle-size distribution function (psd) on $I_{\mathrm{sf}}(q)$ for the NM model and for $K_{\mathrm{s}} > 0$ and $K_{\mathrm{s}} < 0$ (log-log scales). (a, b)~Near saturation ($B_0 = 10 \, \mathrm{T}$) and (c, d)~at remanence ($B_0 = 0 \, \mathrm{T}$). The vertical dashed lines mark the first two minima of the form factor of a homogeneous sphere with a diameter of $D = 2R = 8 \, \mathrm{nm}$ ($I_{\mathrm{sf}}(q) \sim [j_1(qR)/(qR)]^2$ with a first zero at $q_1^{\mathrm{sat}} \cong 9/D = 1.125 \;\mathrm{nm}^{-1}$ and a second zero at $q_2^{\mathrm{sat}} \cong 15.5/D = 1.9375 \;\mathrm{nm}^{-1}$). The mean particle diameter is $\mu = 8 \, \mathrm{nm}$ and the width $\sigma$ of the distribution is varied (see inset).}
\label{fig8}
\end{figure}

In Figs.~\ref{fig5} to \ref{fig8} we present the results for the two-dimensional $d \Sigma_{{\mathrm{sf}}} / d\Omega$ and for the azimuthally-averaged sf SANS cross sections $I_{\mathrm{sf}}(q)$; Figs.~\ref{fig5} and \ref{fig6} present the results for a monodisperse particle ensemble ($D = 8$ nm), Fig.~\ref{fig7} compares the $I_{\mathrm{sf}}(q)$ of individual particles to the randomly-averaged $I_{\mathrm{sf}}(q)$, and Fig.~\ref{fig8} shows the effect of polydispersity on $I_{\mathrm{sf}}(q)$. Near saturation at $B_0 = 10 \, \mathrm{T}$ (upper row in Fig.~\ref{fig5}), we observe for all surface anisotropy models the characteristic $\sin^2\theta \cos^2\theta$~angular anisotropy of $d \Sigma_{{\mathrm{sf}}} / d\Omega$ due to longitudinal magnetization fluctuations [compare Eq.~(\ref{eq:equation1})]. At remanence ($B_0 = 0 \, \mathrm{T}$, middle row), this anisotropy is still prevalent in all the scattering images, but, depending on the surface anisotropy model and the sign of $K_{\mathrm{s}}$, we observe some variations in its overall shape. Compared to the data at $10 \, \mathrm{T}$, we now find large scattering contributions along the horizontal and vertical directions on the detector (except for the CM model). When the external field is further reduced to the respective coercive field (characterized by a zero net magnetization of the particle ensemble, lower row), we see that the anisotropy of the pattern exhibits maxima along the horizontal field direction, except for the CM~case with $K_{\mathrm{s}} < 0$ [Fig.~\ref{fig5}(n)]. We also emphasize that the randomly-averaged magnetic SANS cross section of a particle ensemble is generally anisotropic, even at remanence or in the demagnetized state~\cite{michaeliucrj2023}.

The azimuthally-averaged $I_{\mathrm{sf}}(q)$ near saturation [Fig.~\ref{fig6}(a) and (d)] surprisingly reveal a shift of the form-factor minima to both larger and smaller momentum transfer $q$:~for $K_{\mathrm{s}} > 0$ we observe a shift to larger $q$, while for $K_{\mathrm{s}} < 0$ we see a shift to smaller $q$; these trends remain present in the data also at lower fields. For the case of a random surface anisotropy (RM), this shifting feature is weak. Naively, one may argue that the very slight deviation from a uniform magnetization due to the presence of surface spin disorder [even at $10 \, \mathrm{T}$, compare Fig.~\ref{fig3}(a)$-$(d)] gives rise to a smaller effective ``magnetic particle size'' and a concomitant shift of the minima in $I_{\mathrm{sf}}(q)$ to larger $q$. Since this picture does not become visible in our data we cannot argue that an arbitrary inhomogeneous spin structure generally results in a shift of the extrema to {\it larger} scattering vectors. The origin of this observation remains unknown to us. Moreover, we see in Fig.~\ref{fig6} that the form-factor oscillations for our unimodal particle system are damped in the presence of a strong surface anisotropy. Ignoring the shift of the minima, this behavior somehow mimics the effect of instrumental smearing or of a particle-size distribution function; in other words, the different spin structures [which result from different energies in Eq.~(\ref{eq:Ham-MSP})] result in an intrinsic smearing effect in the sf SANS cross section, even for particles of the same size and shape (compare Fig.~\ref{fig7} below). The difference between positive and negative $K_{\mathrm{s}}$~values becomes most pronounced at remanence and at the coercive fields [Fig.~\ref{fig6}(b), (c), (e), (f)]. For tangential-like spin structures ($K_{\mathrm{s}} < 0$), both the NM and CM models result in a more peak-like functional dependence of $I_{\mathrm{sf}}(q)$ around the first maximum [compare, e.g., the dashed red and blue curves in Fig.~\ref{fig6}(e)], whereas for radial-like structures ($K_{\mathrm{s}} > 0$), we see a more shoulder-like behavior of $I_{\mathrm{sf}}(q)$ [compare, e.g., the solid red and blue curves in Fig.~\ref{fig6}(f)].

As discussed in the introduction, the fact that each nanoparticle of an assembly has its own (random) orientation of the magnetic easy axis of magnetization relative to the laboratory coordinate system gives rise to an averaging procedure over the corresponding distribution. This is illustrated in Fig.~\ref{fig7}, which depicts four selected remanent spin structures and their corresponding contribution (weight) to the azimuthally-averaged sf SANS cross section $I_{\mathrm{sf}}(q)$ of an ensemble of $256$ randomly oriented particles. All materials parameters are the same in these simulations, except that the core-anisotropy direction and the corresponding distribution of the N\'{e}el surface anisotropy are different (since the lattice is rotated). As can be seen, the four spin structures in Fig.~\ref{fig7} are substantially different and exhibit different functional $q$~dependence. As discussed in the previous paragraph, even in the absence of a particle-size distribution function, this gives rise to a kind of {\it intrinsic} broadening effect on $I_{\mathrm{sf}}(q)$ (see Fig.~\ref{fig6}).

The additional smearing effect of a lognormal particle-size distribution is summarized in Fig.~\ref{fig8}. At $B_0 = 10 \, \mathrm{T}$ [Fig.~\ref{fig8}(a) and (b)], we observe the expected behavior, namely, that the form-factor oscillations become progressively damped and washed out. Switching the sign of $K_{\mathrm{s}}$ does (at $10 \, \mathrm{T}$) not change this observation, except that the minima shift into different directions on the $q$~axis (as discussed previously, see Fig.~\ref{fig6}). However, at zero field [Fig.~\ref{fig8}(c) and (d)], the change of the sign of $K_{\mathrm{s}}$ becomes noticeable in $I_{\mathrm{sf}}(q)$. For $K_{\mathrm{s}} > 0$ (radial-like spin structures), the polydispersidy brings no significant effect, since the $I_{\mathrm{sf}}(q)$~curves are already fully damped in the monodisperse case due to the N\'{e}el surface anisotropy (compare Fig.~\ref{fig6}), while for $K_{\mathrm{s}} < 0$ (tangential-like structures) we again observe the usual smearing behavior. These findings are qualitatively similar for the CM model (data not shown).

\subsection{How to analyze the magnetic SANS cross section?}
\label{sec3b}

Quite frequently, azimuthally-averaged magnetic SANS data (or sector averages along the horizontal or vertical directions on the detector) are analyzed by decomposing the cross section into a set of noninterfering spheres or core-shell particles. This represents a purely structural/geometrical approach that is not adapted to the inhomogeneous spin microstrucure of nanoparticles; in other words, there is no physical model behind this procedure. We have also tried to fit the simulation data in Fig.~\ref{fig6} to the sphere and core-shell form factors with and without a distribution of sizes. While some of the saturated data could obviously be satisfactorily described in this way, the data at remanence and at the coercive field (where the largest spin deviations occur) could not be fitted to these models using realistic parameters.

\begin{figure}[tb!]
\centering
\resizebox{1.0\columnwidth}{!}{\includegraphics{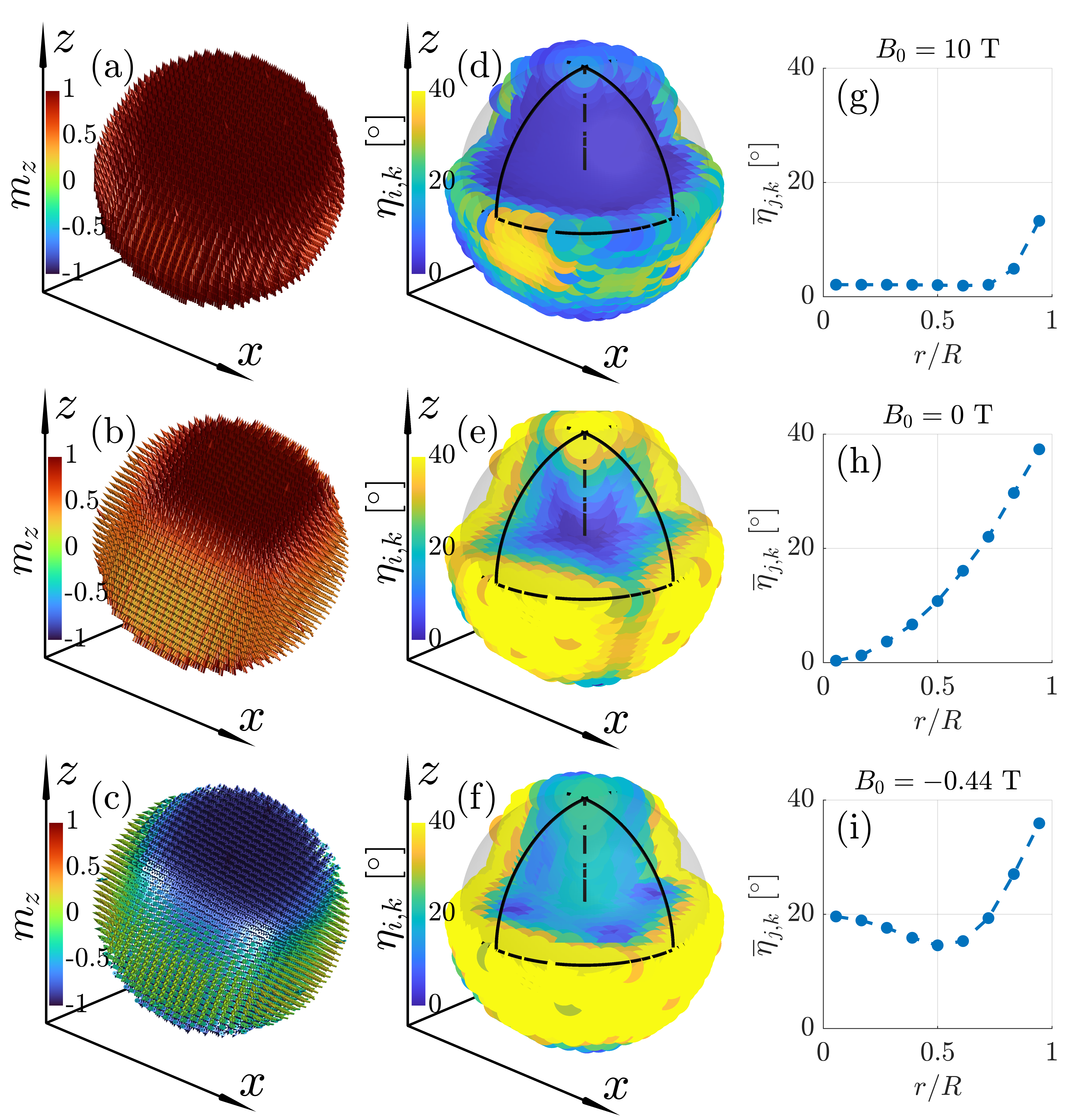}}
\caption{Results for the shell-averaged deviation angle of an individual nanoparticle (NM model with $K_{\mathrm{s}} > 0$, $D = 8 \, \mathrm{nm}$). (a)$-$(c) Real-space spin structures at $B_0 = 10 \, \mathrm{T}$, remanence ($B_0 = 0 \, \mathrm{T}$), and at the coercive field (see insets); (d)$-$(f) the corresponding deviation angles $\eta_{i,k}$ [Eq.~\eqref{eq:LocalSpinDeviationAngle}]; (g)$-$(i) the radial profiles of the shell-averaged deviation angles $\overline{\eta}_{j,k}$ [Eq.~\eqref{eq:ShellAveragedSpinDeviationAngle}].}
\label{fig9}
\end{figure}

To demonstrate that core-shell-type models cannot account for the existing spin inhomogeneity (at low fields) within nanoparticles exhibiting surface anisotropy, we have plotted in Fig.~\ref{fig9} the average spin-misalignment angle as a function of the applied field. This figure requires some explanation that we are providing in the following. Denoting by ``$i$'' the atomic spin index and by ``$k$'' the particle index, we introduce the following quantities:
\begin{align}
\overline{\mathbf{m}}_k &= \frac{1}{\mathcal{N}}\sum_{i=1}^{\mathcal{N}} \mathbf{m}_{i,k} ,
\\
\eta_{i,k} &= \operatorname{arccos}\left(
\frac{\mathbf{m}_{i,k} \cdot \overline{\mathbf{m}}_k}{\|\overline{\mathbf{m}}_k\|} \right) ,
\label{eq:LocalSpinDeviationAngle}
\\
\rho_{i,k} &= \frac{\|\mathbf{r}_{i,k}\|}{R},
\\
\Delta_j &= \left\{ \rho_{i,k}\in \mathbb{R} \; \Big| \; \frac{j-1}{\mathcal{J}} \le \rho_{i,k} < \frac{j}{\mathcal{J}} , \; j = 1,2,...,\mathcal{J} \right\} ,
\\
\overline{\eta}_{j,k} &= \frac{1}{\mathcal{N}_j}\sum_{\rho_{i,k}\in \Delta_j}\eta_{i,k} ,
\label{eq:ShellAveragedSpinDeviationAngle}
\\
\overline{\eta}_{j} &= \frac{1}{\mathcal{K}}\sum_{k=1}^{\mathcal{K}} \overline{\eta}_{j,k}
\label{eq:ShellAveragedSpinDeviationAngleAveragedOverEnsemble}.
\end{align}
At a given external field $B_0$, the vector $\overline{\mathbf{m}}_k$ represents the average magnetization of particle $k$, which contains a total of $\mathcal{N}$ spins (the same for all particles, since we consider only the monodisperse case). The quantity $\eta_{i,k}$ denotes the angle of deviation between the local spin $i$ in particle $k$ relative to the unit vector along the average magnetization of particle $k$. $\rho_{i,k}$ is the normalized radial distance from the origin (in particle $k$) to the spin $i$, and $\Delta_j$ describes the decomposition of a given spherical particle (with radius $R$) into a total number of $\mathcal{J}$ shells (intervals) with a shell thickness of $1/\mathcal{J}$. $\overline{\eta}_{j,k}$ is the shell-averaged deviation angle of particle $k$, where $\mathcal{N}_j$ denotes the number of spins within the $j$th shell of the $k$th particle. Finally, $\overline{\eta}_{j}$ [Eq.~\eqref{eq:ShellAveragedSpinDeviationAngleAveragedOverEnsemble}] denotes the shell-averaged deviation angle averaged over the ensemble of $\mathcal{K}$ particles. The latter quantity is also displayed in the Supplemental Material of this paper~\cite{michael2024sm}. For the following discussion, the average deviation angle $\overline{\eta}_{j,k}$, which can be interpreted as a measure for the spin inhomogeneity of particle $k$, is of relevance; $\overline{\eta}_{j,k} = 0^{\circ}$ corresponds to the uniform single-domain state.

Figure~\ref{fig9} shows the results for the shell-averaged deviation angle $\overline{\eta}_{j,k}$ of an individual nanoparticle (for the NM model with $K_{\mathrm{s}} > 0$). The results for $\overline{\eta}_{j,k}$ for the other surface anisotropy models are similar (data not shown). Displayed are real-space spin structures for an individual particle near saturation, at remanence, and at the (near) coercive field [Fig.~\ref{fig9}(a), (b), (c)]; the subpanels (d), (e), (f) show the data for $\eta_{i,k}$ [Eq.~\eqref{eq:LocalSpinDeviationAngle}] and the subpanels (g), (h), (i) display $\overline{\eta}_{j,k}$ [Eq.~\eqref{eq:ShellAveragedSpinDeviationAngle}]. While the $\eta_{i,k}$ represent the three-dimensional distribution of the {\it local} spin deviation angles in the unit sphere, the shell-averaged spin deviation angles $\overline{\eta}_{j,k}$ are of highest relevance for the following discussion.

In the case close to saturation [$B_0 = 10 \, \mathrm{T}$, Fig.~\ref{fig9}(g)], the $\overline{\eta}_{j,k}$ indicate an approximate core-shell-like behavior, where the $\overline{\eta}_{j,k} \cong 0^{\circ}$ are constant inside the core of the nanoparticle (up to $r/R \cong 0.75$), and then increase with a parabolic functional dependence towards the particle surface. At lower fields, the spin inhomogeneities spread towards the center of the nanoparticle, such that we cannot discern anymore a sharp core-shell-like transition~\cite{michaelphysscr2023}. In fact, at remanence [Fig.~\ref{fig9}(h)] we find a parabolic $\overline{\eta}_{j,k}(r/R)$~dependence that extends over the whole particle, and even becomes more nonlinear at the coercive field [Fig.~\ref{fig9}(i)]. These results then suggest that it is not possible/permissible to fit such low-field data to a set of noninterfering core-shell particles, which represents a structural model that is not adapted to the nonuniform three-dimensional spin distribution within nanoparticles.

\begin{figure}[tb!]
\centering
\resizebox{0.90\columnwidth}{!}{\includegraphics{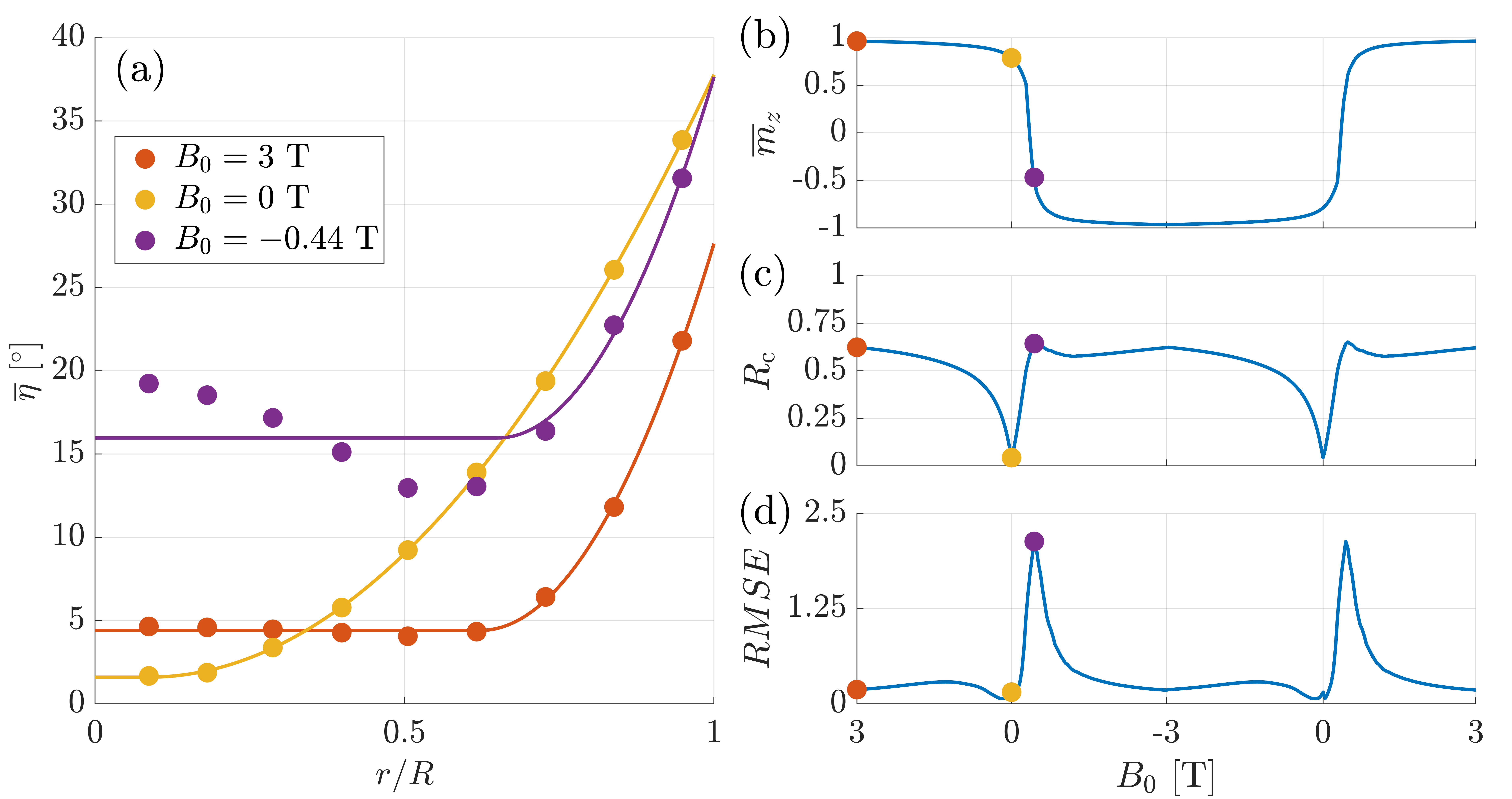}}
\caption{Results of the fit analysis for $\overline{\eta}(\xi)$, the shell-averaged deviation angle averaged over the ensemble of $\mathcal{K}$ particles [Eq.~\eqref{eq:ShellAveragedSpinDeviationAngleAveragedOverEnsemble}] (for the NM case with $K_{\mathrm{s}} < 0$). (a)~Simulation data for $\overline{\eta}(\xi = r/R)$ (dots) are shown for three selected applied fields $B_0 = \{ 3, \;0,\; -0.44 \}$~T and are fitted to Eq.~\eqref{eq:ModelFunction_eta} (solid lines). It is seen that at the coercive field (regime of magnetization reversal), the second-order model Eq.~\eqref{eq:ModelFunction_eta} fails. (b)~Average magnetization $\overline{m}_z(B_0)$ along the hysteresis loop. (c) and (d)~Field dependence of $R_{\mathrm{c}}$ and of the root-mean-square error (RMSE) between the model fit and the simulation data. The dots in (b)$-$(d) correspond to the three cases shown in (a).}
\label{fig10}
\end{figure}

To further illustrate this, we present in Fig.~\ref{fig10} a model fit of the $\overline{\eta}$~profiles [Eq.~\eqref{eq:ShellAveragedSpinDeviationAngleAveragedOverEnsemble}] along the hysteresis loop. As a model, we use the following piecewise polynomial function:
\begin{align}
\overline{\eta}(\xi = r/R) = 
\begin{cases}
\eta_0 & , \hspace{0.4cm} 0 \le \xi \le R_{\mathrm{c}}
\\
\eta_0 + \eta_1 (\xi - R_{\mathrm{c}})^2 & , \hspace{0.4cm} R_{\mathrm{c}} \le \xi \le 1
\end{cases}
\label{eq:ModelFunction_eta} ,
\end{align}
which consists of a constant part with magnitude $\eta_0$ (defined for $\xi \le R_{\mathrm{c}}$) and a parabolic part with prefactor $\eta_1$ (defined for $\xi \ge R_{\mathrm{c}}$). $R_{\mathrm{c}}$ is the reduced core radius, which reflects the idea of a core-shell nanoparticle. It can be seen in Fig.~\ref{fig10} that in most cases Eq.~\eqref{eq:ModelFunction_eta} describes the averaged radial behavior of the system, except for the regime of magnetization reversal (finite interval around the coercive field), where a description using a higher-order polynomial (degree $>2$) might be more appropriate. We emphasize that the reduced core radius $R_{\mathrm{c}}$ is continuously changing with the field [Fig.~\ref{fig10}(c)], similar to what has been reported by Z\'akutn\'a~et~al.~\cite{zakutna2020}. This observation in conjunction with the fact that the simulation data [Fig.~\ref{fig10}(a)] do not exhibit a step-function profile (except approximately at the highest field) demonstrates that a core-shell-type model is not suitable for the analysis of the corresponding scattering data.

In the following, we develop a novel power-series magnetization vector field model that provides an analytical expression for the azimuthally-averaged sf SANS cross section $I_{\mathrm{sf}}(q)$ of nanoparticles (and any magnetic SANS cross section in general). This approach takes into account arbitrary spin inhomogeneity and particle shape and is not necessarily restricted to the presence of surface anisotropy as the main mechanism to generate intraparticle spin disorder.

\section{Multi-particle power-series analysis of the magnetic SANS cross section}
\label{sec4}

In the discussion of magnetic SANS from nanoparticles, several features need to be distinguished: (i)~the particle-size distribution (including the particle-shape distribution), (ii)~the spatial distribution of the particles within the sample, and (iii)~the total magnetization vector field of the sample (including all particles). Based on an analytical calculation of the magnetic SANS cross section from nanoparticles with N\'{e}el surface anisotropy~\cite{adamsjacana2022}, which provided an explicit expression for the two-dimensional magnetic SANS cross section beyond the superspin model, we introduce here a power-series analysis of a multi-particle system taking into account arbitrary intrinsic magnetization distributions. As shown in Appendix~\ref{appendixa}, by using this method we are able to derive an analytical expression for $I_{\mathrm{sf}}(q)$ up to the second-order spatial dependence of the magnetization vector field $\mathbf{M}(\mathbf{r})$. The expression for $I_{\mathrm{sf}}(q)$ may be used for the analysis of experimental neutron data. By way of illustration, we use our power-series model (up to the second order) to fit the SANS results from the atomistic simulations of nanoparticles with different types of surface anisotropy as well as from micromagnetic continuum simulations of larger nanoparticles with inherent vortex-type spin configurations~\cite{evelynprb2023}.

The final result for the azimuthally-averaged sf SANS cross section $I_{\mathrm{sf}}(q)$ for the perpendicular scattering geometry is given by (see Appendix~\ref{appendixa}):
\begin{align}
I_{\mathrm{sf}}(q) = \sum_{k=0}^{6} I_{\mathrm{sf}}^k \, g_k(qR),
\label{eq:AzimuthallyAveragedSANScrosssection_SphericalParticlesSecondOrdermaintext}
\end{align}
where the $I_{\mathrm{sf}}^k$ are constant prefactors, and the radially-symmetric functions $g_k(qR)$ are given by Eqs.~(\ref{eq:MagnetizationModel_g0})$-$(\ref{eq:MagnetizationModel_g6}). In the perfectly saturated state, the higher-order coefficients in Eq.~\eqref{eq:AzimuthallyAveragedSANScrosssection_SphericalParticlesSecondOrdermaintext} vanish and the remaining zeroth-order term is given by the well-known homogeneous sphere form factor:
\begin{align}
I_{\mathrm{sf}}(q; B_0\rightarrow\infty) = I_{\mathrm{sf}}^{0, \mathrm{sat}}  g_0(qR) = I_{\mathrm{sf}}^{0, \mathrm{sat}} \left[ \frac{\sin(qR) -  qR \cos(qR)}{(q R)^3} \right]^2 .
\label{satresult}
\end{align}
Equation~\eqref{eq:AzimuthallyAveragedSANScrosssection_SphericalParticlesSecondOrdermaintext} is one of the central results of this paper. It represents an easy-to-use fit function for azimuthally-averaged magnetic SANS cross sections of ensembles of monodisperse and dilute spherical particles with up to 8 free fit parameters ($R$ and $I_{\mathrm{sf}}^k,\,k=0\ldots 6$); $I_{\mathrm{sf}}(q)$ depends linearly on $I_{\mathrm{sf}}^k$, but nonlinearly on the sphere radius $R$. We emphasize that although Eq.~\eqref{eq:AzimuthallyAveragedSANScrosssection_SphericalParticlesSecondOrdermaintext} has been derived for the purely magnetic sf SANS cross section, it is equally well applicable to the purely magnetic SANS cross section that might be obtained by means of unpolarized SANS measurements:~as shown e.g.\ in Refs.~\cite{bersweiler2019,michelsbook}, subtracting the nuclear and magnetic unpolarized SANS cross section at saturation from the nuclear and magnetic unpolarized SANS at a lower field (assuming a field-independent nuclear scattering) results in a purely magnetic (difference) SANS cross section that is closely related to the sf SANS (just a different combination of the magnetic Fourier components). Likewise, Eq.~\eqref{eq:AzimuthallyAveragedSANScrosssection_SphericalParticlesSecondOrdermaintext} is also applicable to any magnetic SANS cross section measured in the parallel scattering geometry. The coefficients $I_{\mathrm{sf}}^k$ may in these cases simply take on different values.

As dicussed in Appendix~\ref{appendixa}, the coefficients $I_{\mathrm{sf}}^k$ may generally depend on the temperature, the applied magnetic field, on the magnetic interactions (e.g., symmetric and antisymmtric exchange, magnetic anisotropy, magnetodipolar interaction), and in particular on the radius $R$ of the nanoparticle. Therefore, in the presence of a particle-size distribution function $w(R)$, the $I_{\mathrm{sf}}^k$ become functions of $R$. One may then either assume certain distribution functions for the $I_{\mathrm{sf}}^k$ (e.g., Gaussian), but this would lead to an unreasonably large number of free fitting parameters. Instead, a more practical approach is to carry out a fitting procedure over $w(R)$ and to interpret the $I_{\mathrm{sf}}^k$ as ensemble-averaged quantities, which then implies that they are uniformly distributed over the particle sizes.

\begin{figure}[tb!]
\centering
\resizebox{0.90\columnwidth}{!}{\includegraphics{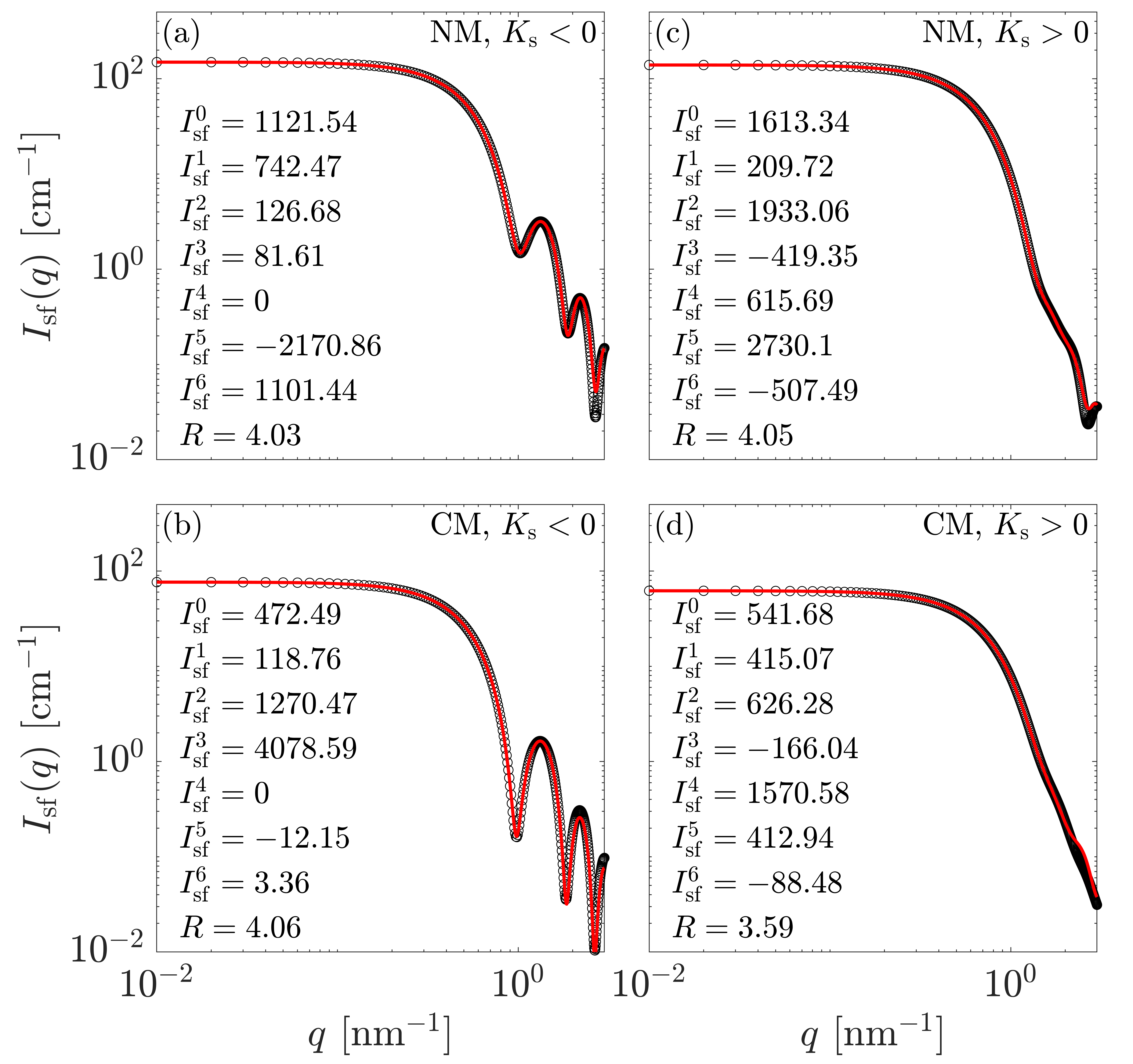}}
\caption{Fits of the model function Eq.~\eqref{eq:AzimuthallyAveragedSANScrosssection_SphericalParticlesSecondOrdermaintext} (solid lines) to the sf simulation results of monodisperse ensembles of spherical nanoparticles with different types of surface anisotropy ($\circ$) ($D = 8 \, \mathrm{nm}$, $B_0 = 0 \, \mathrm{T}$). The coefficients $I_{\mathrm{sf}}^{k}$ are given in units of cm$^{-1}$ and the radius $R$ is given in units of nm. In panels~(a,b) we show the cases with $K_{\mathrm{s}} < 0$ and in (c,d) we display the results with $K_{\mathrm{s}} > 0$. For the fit analysis, we have used the Levenberg-Marquardt algorithm.}
\label{fig11}
\end{figure}

To verify that Eq.~\eqref{eq:AzimuthallyAveragedSANScrosssection_SphericalParticlesSecondOrdermaintext}, which is based on a second-order polynomial expansion of the magnetization vector field, may be used to explain different features in the magnetic SANS cross section, we have fitted the free parameters in Eq.~\eqref{eq:AzimuthallyAveragedSANScrosssection_SphericalParticlesSecondOrdermaintext} to several selected simulation data. Figure~\ref{fig11} displays the atomistic simulation results for $I_{\mathrm{sf}}(q)$ for the NM and CM models at remanence (open circles) along with the fits to Eq.~\eqref{eq:AzimuthallyAveragedSANScrosssection_SphericalParticlesSecondOrdermaintext} (solid lines). It can be seen that the second-order model [Eq.~\eqref{eq:AzimuthallyAveragedSANScrosssection_SphericalParticlesSecondOrdermaintext}] describes the sf data very well; some small deviations occur at the largest momentum transfers $q$. As predicted, the coefficients $I_{\mathrm{sf}}^{0}, \; I_{\mathrm{sf}}^{1} , \; I_{\mathrm{sf}}^{2}, \; I_{\mathrm{sf}}^{4}$ are all positive, while the $I_{\mathrm{sf}}^{3}, \; I_{\mathrm{sf}}^{5}, \; I_{\mathrm{sf}}^{6}$ can take on positive as well as negative values. The radii $R$ are all very close to $4 \, \mathrm{nm}$, except for the CM model with $K_{\mathrm{s}} > 0$, where the numerical minimization algorithm finds $R \cong 3.6 \, \mathrm{nm}$, which is significantly smaller than the true geometrical size. This result might be related to the featureless $I_{\mathrm{sf}}(q)$~curve for this case. By switching the sign of $K_{\mathrm{s}}$ we see for both NM and CM models that the signs of the coefficients $I_{\mathrm{sf}}^{3},\; I_{\mathrm{sf}}^{5}, \; I_{\mathrm{sf}}^{6}$ are switched (compare Appendix~\ref{appendixa}). Furthermore, we note that for the cases $K_{\mathrm{s}} < 0$ the coefficient $I_{\mathrm{sf}}^{4}$ is numerically equal to zero (approaching zero from above).

\begin{figure}[tb!]
\centering
\resizebox{1.00\columnwidth}{!}{\includegraphics{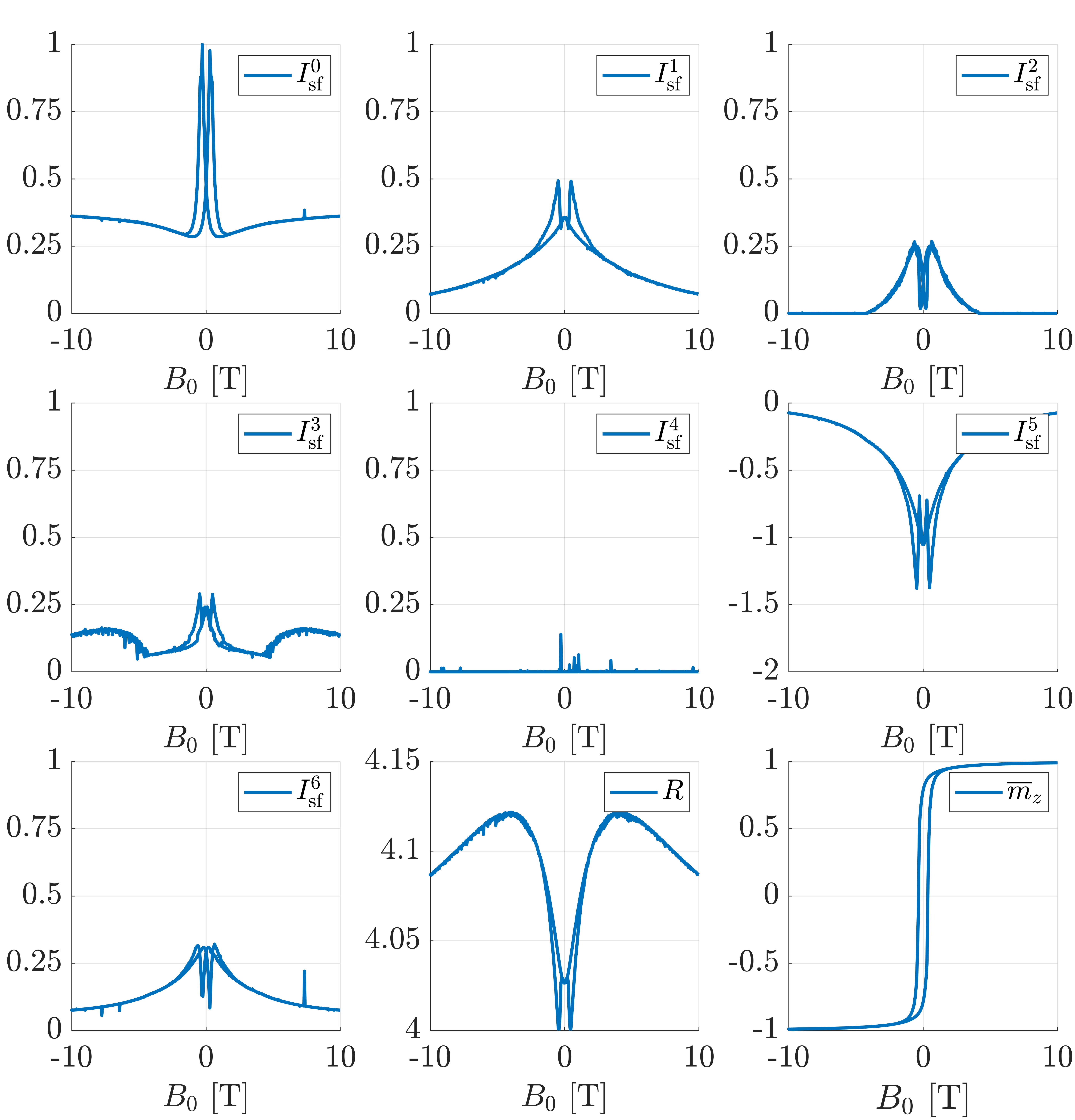}}
\caption{Field evolution of the fitting parameters $I_{\mathrm{sf}}^{k}$ and $R$ in Eq.~\eqref{eq:AzimuthallyAveragedSANScrosssection_SphericalParticlesSecondOrdermaintext}. Spin flip data of the NM model with $K_{\mathrm{s}} < 0$ and $D = 8 \, \mathrm{nm}$ are analyzed. The lowest right panel shows the corresponding normalized magnetization curve $\overline{m}_z(B_0)$. The coefficients $I_{\mathrm{sf}}^k$ are normalized to the maximum value of $I_{\mathrm{sf}}^0$, the particle radius $R$ is in units of nanometers.}
\label{fig12}
\end{figure}

The field dependence of the fitting parameters for the NM model with $K_{\mathrm{s}} < 0$ is presented in Fig.~\ref{fig12}. It can be seen that the coefficients $I_{\mathrm{sf}}^{k}$ and the radius $R$ (with the exception of $I_{\mathrm{sf}}^4$) behave approximately mirror symmetrically with respect to positive and negative applied fields. For increasing field strength ($|B_0|\rightarrow \infty$), the parameters $I_{\mathrm{sf}}^{1}$ to $I_{\mathrm{sf}}^{6}$ tend to zero (as expected), indicating that the internal magnetization structure becomes progressively more uniform, approaching the saturated case that is given by $I_{\mathrm{sf}}(q) = I_{\mathrm{sf}}^{0, \mathrm{sat}} g_0(qR)$ [Eq.~(\ref{satresult})]. Note the apparent field dependence of the fit value for $R$, which is a consequence of the numerical fit with the second-order model (the fluctuation in $R$ is in the 1~\AA~regime). The results of fitting for the field dependencies of the coefficients $I_{\mathrm{sf}}^{k}$ and for the radii $R$ are qualitatively similar for the other surface anisotropy models.

\begin{figure}[tb!]
\centering
\resizebox{1.0\columnwidth}{!}{\includegraphics{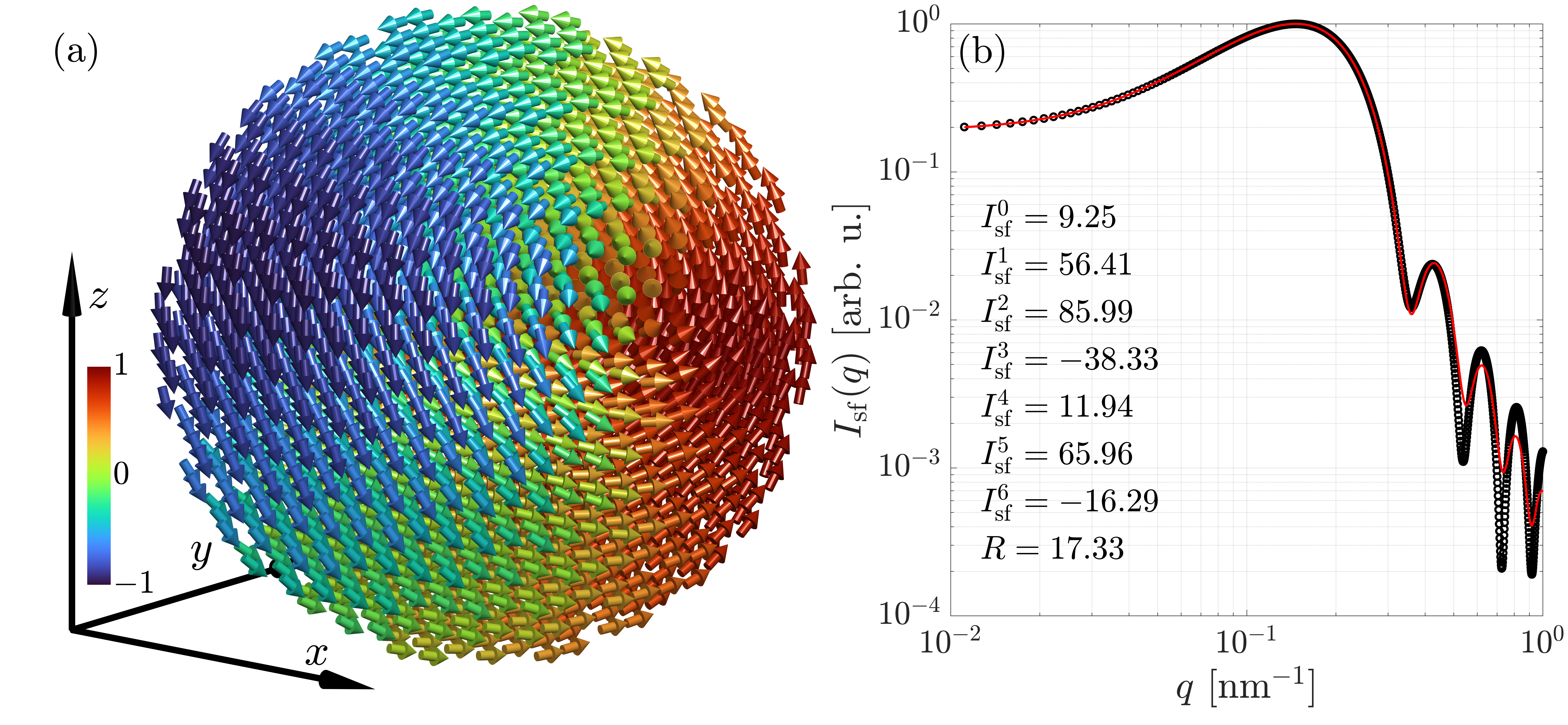}}
\caption{(a)~Vortex spin structure in a single spherical Fe nanoparticle at remanence ($B_0 = 0 \, \mathrm{T}$, $D = 34 \, \mathrm{nm}$). The particle has previously been saturated along the $z$~direction. (b)~($\circ$)~Azimuthally-averaged sf SANS cross section $I_{\mathrm{sf}}(q)$ of a corresponding ensemble of randomly-oriented particles and fit to Eq.~\eqref{eq:AzimuthallyAveragedSANScrosssection_SphericalParticlesSecondOrdermaintext} (solid line) (log-log scale).} 
\label{fig13}
\end{figure}

Atomistic simulations, in particular of systems with larger particle sizes, do not allow for the inclusion of the magnetodipolar interaction, which is due to the related high numerical cost~\cite{koehlerjac2021}. Recent micromagnetic simulations have shown that the dipolar energy is responsible for the formation of vortex-type spin structures in spherical Fe nanoparticles~\cite{laura2020,evelynprb2023}. Figure~\ref{fig13}(a) shows as an example a vortex structure in a $34 \, \mathrm{nm}$-sized cubic Fe particle at remanence. The typical discretization volume in such micromagnetic computations is $2 \times 2 \times 2 \, \mathrm{nm}$, which allows one to compute, in a reasonable time, the randomly-averaged magnetic SANS cross section of particles with sizes up to a few hundreds of nanometers. The results in Fig.~\ref{fig13}(b) demonstrate that the low-$q$ part of the simulated scattering curve is very well described by the analytical model [Eq.~\eqref{eq:AzimuthallyAveragedSANScrosssection_SphericalParticlesSecondOrdermaintext}], but that deviations occur at larger $q$. While the positions of the extrema at the large $q$ are reproduced, their fine details are not. The value for the particle size is recovered correctly. The vortex structure in Fig.~\ref{fig13} is an example of a highly inhomogeneous spin texture. Even for this situation, our model is able to provide a reasonable description of the main features.

\section{Conclusions and Outlook}
\label{sec5}

The signature of surface anisotropy in magnetic nanoparticles in their spin-flip (sf) small-angle neutron scattering (SANS) cross section has been investigated by means of atomistic simulations. Taking into account the isotropic exchange interaction, an external magnetic field, a uniaxial or cubic magnetic core anisotropy, and various models for the surface anisotropy (N\'{e}el, conventional, random), we have computed the sf SANS cross section from the obtained equilibrium spin structures using the Landau-Lifshitz equation of motion. The sign of the surface anisotropy constant $K_{\mathrm{s}}$ is related to the appearance of tangential-like ($K_{\mathrm{s}} < 0$) or radial-like ($K_{\mathrm{s}} > 0$) spin textures. These can be distinguished in the azimuthally-averaged sf signal via their dependence on the momentum-transfer vector $q$. The scattering data cannot be described by the well-known and often-used analytical expressions for uniformly magnetized spherical or core-shell particles, in particular at remanence or at the coercive field. Even if all the particles have the same size and shape, their spin structures are generally different due to the fact that their anisotropy axes are differently oriented with respect to the external magnetic field. This gives rise to a kind of intrinsic spin-structure-related smearing effect in the SANS cross section. Inspired by these facts, and based on a second-order power-series expansion of the magnetization vector field, we have developed a novel and easy-to-implement minimal model for the azimuthally-averaged magnetic SANS cross section [Eq.~(\ref{eq:AzimuthallyAveragedSANScrosssection_SphericalParticlesSecondOrdermaintext})]. We emphasize that the theory is valid for an arbitrary spin inhomogeneity and is not restricted to the specific case of surface anisotropy. It has been shown that Eq.~(\ref{eq:AzimuthallyAveragedSANScrosssection_SphericalParticlesSecondOrdermaintext}) describes very well our simulation data as well as more complex spin patterns such as vortex-like structures. In this way, it has become possible to describe the behavior of a very large number of atomic spins ($11363$~spins in an $8 \, \mathrm{nm}$-sized particle times $256$ different easy-axis orientations) by only seven expansion coefficients $I_{\mathrm{sf}}^k$ and some basis functions $g_k(qR)$.

Regarding future studies, of course, one could include high-order terms in the power-law expansion for the magnetization [Eq.~(\ref{eq:MagnetizationPowerExpansion})] or a particle-size distribution function. However, this would significantly increase the number of free parameters in the model ($I_{\mathrm{sf}}^k$) and would very likely not provide further insights into the problem. Rather, one should focus on the physical interpretation of the $I_{\mathrm{sf}}^k$ within the second-order approach. For instance, one could systematically analyze the field dependence of the $I_{\mathrm{sf}}^k$ for the different signs and strengths of the surface anisotropy constants $K_{\mathrm{s}}$, study their behavior for different $K_{\mathrm{s}} / K_{\mathrm{c}}$ and $K_{\mathrm{s}} / K_{\mathrm{u}}$~ratios (here, we use a rather large value of $K_{\mathrm{s}} / K_{\mathrm{c}} \cong 9206$), or one could implement more realistic lattice structures (e.g., of spinel type) with complex exchange interactions. Another interesting study would be the comparison of the outcome of atomistic and coarse-grained micromagnetic computations for the sf SANS cross section. From the sample synthesis point of view, it would be desirable to prepare oriented nanoparticle assemblies, where the magnetic easy axes of the particles all point into the same direction. This in conjunction with a uniform particle-size distribution will significantly facilitate the scattering-data analysis since the corresponding averages over these features can be straightforwardly carried out.

We also refer the reader to the Supplemental Material~\cite{michael2024sm} of this paper, where several videos are provided that show the magnetization curve, real-space spin structure, particle-ensemble-averaged deviation angle, as well as the 2D and 1D sf SANS cross sections during the magnetization-reversal process. These quantities are shown for different sign combinations of the cubic/uniaxial core and surface anisotropy constants. Additionally, we show a movie that, starting from a single nanoparticle, highlights the stepwise built-up of the randomly-averaged sf SANS cross section corresponding to a total of $256$ particles with different (random) orientations of both the core-anisotropy axes and the related surface anisotropies.

\acknowledgments{Michael Adams, Evelyn Pratami Sinaga, and Andreas Michels thank the National Research Fund of Luxembourg for financial support (AFR Grant No.~15639149 and PRIDE MASSENA Grant).}

\appendix

\section{Multi-particle power-series analysis of the magnetic SANS cross section}
\label{appendixa}

\subsection{Magnetization power-series expansion, Fourier cross-correlation matrix, and magnetic SANS cross section}

\begin{figure}[tb!]
\centering
\resizebox{0.50\columnwidth}{!}{\includegraphics{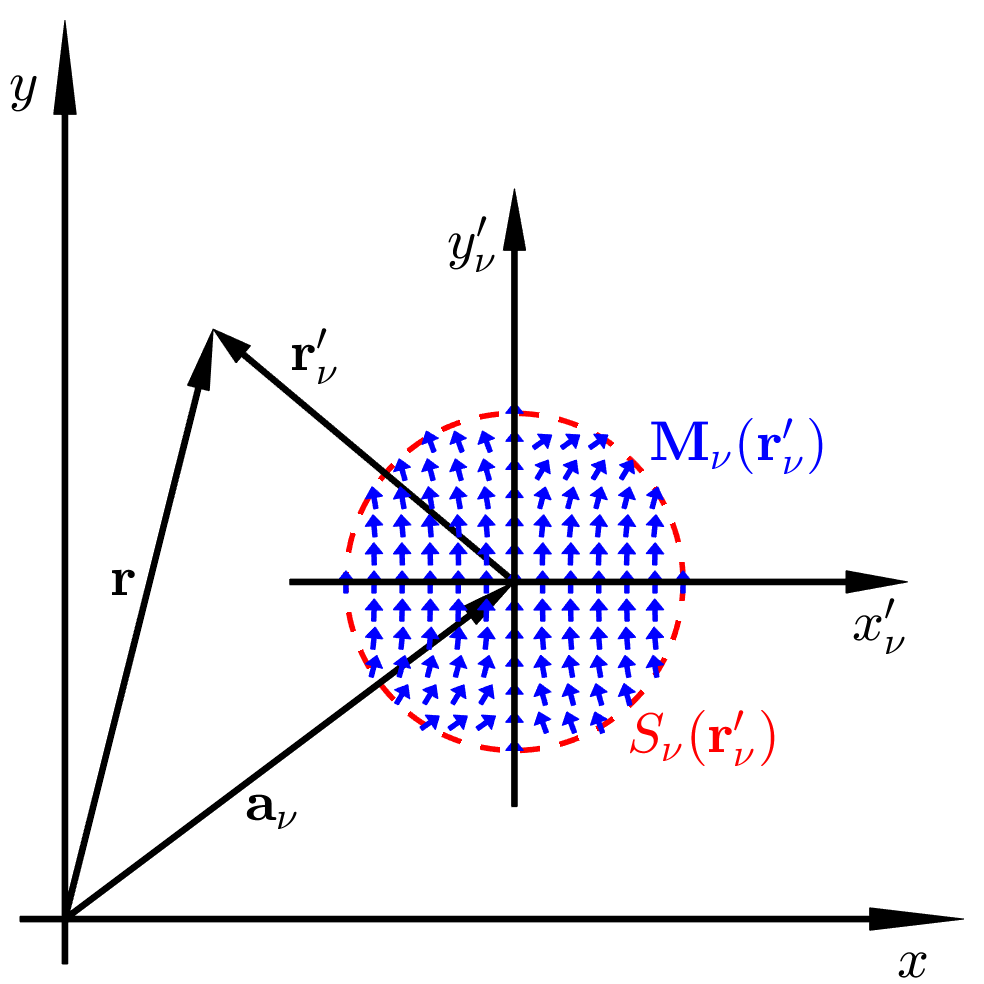}}
\caption{Sketch illustrating the relationship between the global unprimed ($\mathbf{r}$) laboratory coordinate system and the local primed ($\mathbf{r}_{\nu}'$) system of particle $\nu$ with magnetization $\mathbf{M}_{\nu}(\mathbf{r}_{\nu}')$ and shape function $S_{\nu}(\mathbf{r}_{\nu}')$. $\mathbf{a}_{\nu}$ is a constant shift vector that points from the origin of the global $\mathbf{r}$~coordinate system to the origin of the local $\mathbf{r}_{\nu}'$~system. For simplicity, the $z$~coordinate specifying the third space dimension has been ignored.}
\label{fig14}
\end{figure}

We consider an ensemble of magnetic nanoparticles rigidly embedded in a nonmagnetic and homogeneous matrix. The global magnetization vector field of the system, $\mathbf{M}(\mathbf{r}) = \{ M^{x}(\mathbf{r}), M^{y}(\mathbf{r}), M^{z}(\mathbf{r}) \}$, is generally a discontinuous function, since $\mathbf{M}$ vanishes in the space between the particles; $\mathbf{r}=\{x,y,z\}$ is the position vector in the laboratory frame. For the formulation of this discontinuous behavior, we use the indicator function (or particle shape function with particle index $\nu$)
\begin{align}
    S_{\nu}(\mathbf{r}_{\nu}') = \begin{cases}
        1 & , \;  \mathbf{r}_{\nu}'\in V_{\nu}'
        \\
        0 & , \; \mathbf{r}_{\nu}'\notin V_{\nu}'
    \end{cases},
\end{align}
where $V_{\nu}'\subset\mathbb{R}^3$ denotes the set of points within the $\nu$-th particle volume with reference to the local particle frame, and $\mathbf{r}_{\nu}'=\{x_{\nu}', y_{\nu}', z_{\nu}'\}$ represent the local coordinates (see Fig.~\ref{fig14}). The transformation between the global point set $V_{\nu}$ and the local point set $V_{\nu}'$ is then obtained by $V_{\nu}' = \{ \mathbf{r} - \mathbf{a}_{\nu}: \mathbf{r}\in V_{\nu} \}$ (with the inverse transformation: $V_{\nu} = \{ \mathbf{r}_{\nu}' + \mathbf{a}_{\nu}: \mathbf{r}_{\nu}'\in V_{\nu}' \}$), where $\mathbf{a}_{\nu} = \{a_{\nu}^{x}, a_{\nu}^{y}, a_{\nu}^{z} \}$ is a constant shift vector that points from the origin of the global $\mathbf{r}$~coordinate system to the origin of the local $\mathbf{r}_{\nu}'$~system. The corresponding linear coordinate transformation is then given by $\mathbf{r}_{\nu}' = \mathbf{r} - \mathbf{a}_{\nu}$, while the volume $v_{\nu}$ of the $\nu$-th particle is obtained via integration of the corresponding shape function:
\begin{align}
    v_{\nu} = \int_{\mathbb{R}^3} S_{\nu}(\mathbf{r}_{\nu}') d^3r_{\nu}'.
\end{align}
To account for an inhomogeneous magnetic microstructure, we describe the Cartesian magnetization vector field components $M_{\nu}^{\mu}$ (with $\mu\in\{x,y,z\}$) for the $\nu$-th particle as the product of its shape function and a power series:
\begin{align}
    M_{\nu}^{\mu}(\mathbf{r}_{\nu}') = S_{\nu}(\mathbf{r}_{\nu}')
    \sum_{k,m,n=0}^{\infty} M_{\nu,(k,m,n)}^{\mu} x_{\nu}'^{k} y_{\nu}'^{m} z_{\nu}'^{n}, 
\end{align}
where $M_{\nu,(k,m,n)}^{\mu}$ are arbitrary constant expansion coefficients, which may depend on temperature, applied magnetic field, and the type of material. The global Cartesian magnetization vector field components $M^{\mu}$ then follow as the sum over the individual magnetization components $M_{\nu}^{\mu}$ shifted by $\mathbf{a}_{\nu}$: 
\begin{align}
    M^{\mu}(\mathbf{r}) &= \sum_{\nu=1}^{\mathcal{K}}M_{\nu}^{\mu}(\mathbf{r}-\mathbf{a}_{\nu}) 
    \\
    &= 
    \sum_{\nu=1}^{\mathcal{K}} \left[S_{\nu}(\mathbf{r}-\mathbf{a}_{\nu})
    \sum_{k,m,n=0}^{\infty} M_{\nu,(k,m,n)}^{\mu} (x-a_{\nu}^{x})^{k} (y-a_{\nu}^{y})^{m} (z-a_{\nu}^{z})^{n}\right],
\label{eq:GlobalMagnetizationPowerSeries}
\end{align}
$\mathcal{K}$ being the number of particles in the assembly. For the further derivations, we prefer the Einstein and multi-index notation. Using these notation concepts, Eq.~\eqref{eq:GlobalMagnetizationPowerSeries} reads: 
\begin{align}
M^{\mu}(\mathbf{r}) =  S_{\nu}(\mathbf{r}-\mathbf{a}_\nu)M_{\nu,\boldsymbol{\alpha}}^{\mu}  (\mathbf{r}-\mathbf{a}_\nu)^{\boldsymbol{\alpha}},
\label{eq:MagnetizationPowerExpansion}
\end{align}
where $\boldsymbol{\alpha}=(k,m,n)$ represents a multi-index. The zero-order case of $\boldsymbol{\alpha}=(0,0,0)$ corresponds to the situation of an ensemble of uniformly magnetized nanoparticles. Higher-order terms in this series take into account the local spatial nonuniformities in $\mathbf{M}$. 

For the computation of the magnetic SANS cross section, the next step is to perform the spatial Fourier transform
\begin{align}
    \widetilde{M}^\mu(\mathbf{q}) = \frac{1}{(2\pi)^{3/2}}\int_{\mathbb{R}^{3}} M^{\mu}(\mathbf{r}) \exp(-i\mathbf{q}\cdot\mathbf{r}) d^3r .
\end{align}
Instead of direct integration, we can use the shift and derivation theorem of Fourier theory, such that the Fourier transform of Eq.~\eqref{eq:MagnetizationPowerExpansion} can be expressed as
\begin{align}
\widetilde{M}^{\mu}(\mathbf{q}) =  i^{|\boldsymbol{\alpha}|} M_{\nu,\boldsymbol{\alpha}}^{\mu} \exp(-\mathrm{i}\mathbf{q}\cdot\mathbf{a}_{\nu})\partial^{\boldsymbol{\alpha}} \widetilde{S}_{\nu}(\mathbf{q}),
\label{eq:FourierTransformMagnetizationPowerExpansion}
\end{align}
where $i$ is the imaginary number ($i^2=-1$). 

In the sequel, the derivative $\partial^{\boldsymbol{\alpha}}$, with $\boldsymbol{\alpha}=(k,m,n)$, will denote the $|\boldsymbol{\alpha}|$-th order mixed partial derivative 
\begin{align}
\partial^{\boldsymbol{\alpha}} \equiv \frac{\partial^k}{\partial q_x^k}\frac{\partial^m}{\partial q_y^m}\frac{\partial^n}{\partial q_z^n},
\end{align}
with $|\boldsymbol{\alpha}| = k + m + n$ being the sum of components of the multi-index $\boldsymbol{\alpha}=(k,m,n)$. Likewise, the compact sum $\sum_{\boldsymbol{\alpha}}$ should be understood as the triple sum $\sum_k\sum_m\sum_n$. $\widetilde{S}_{\nu}(\mathbf{q})$ is the Fourier transform of the indicator function defined by
\begin{align}
    \widetilde{S}_{\nu}(\mathbf{q}) = \frac{1}{(2\pi)^{3/2}}\int_{\mathbb{R}^{3}} S_{\nu}(\mathbf{r}) \exp(-\mathrm{i}\mathbf{q}\cdot\mathbf{r}) d^3r.
\end{align}
Next, introducing the following Fourier cross-correlation functions $\widetilde{\Gamma}^{\iota\kappa} : \mathbb{R}^3\rightarrow \mathbb{C}$~\footnote{{We emphasize that in the spin-flip SANS cross section $d\Sigma_{\mathrm{sf}}/d\Omega$ [Eq.~(\ref{eq:equation1}) without the chiral function] the combinations of the cross-correlation functions $\widetilde{\Gamma}^{\iota\kappa}$ cancel out their imaginary parts (more specifically, in terms $\widetilde{\Gamma}^{yz} + \widetilde{\Gamma}^{zy} = \widetilde{M}_y \widetilde{M}_z^{\ast} + \widetilde{M}_y^{\ast} \widetilde{M}_z$), which is why only the real-parts of the $\widetilde{\Gamma}^{\iota\kappa}$ are effective.}} with $\iota, \kappa \in \{x,y,z\}$ (``$*$'' stands for the complex conjugate),
\begin{align}
\widetilde{\Gamma}^{\iota\kappa}(\mathbf{q})
&=
\left[\widetilde{M}^{\iota}(\mathbf{q})\right] 
\left[\widetilde{M}^{\kappa}(\mathbf{q})\right]^{\ast}\\
&= 
i^{|\boldsymbol{\alpha}|-|\boldsymbol{\beta}|}
M_{\nu,\boldsymbol{\alpha}}^{\iota}
M_{\mu,\boldsymbol{\beta}}^{\kappa}
\exp(-\mathrm{i}\mathbf{q}\cdot[\mathbf{a}_{\nu} - \mathbf{a}_{\mu}]) \partial^{\boldsymbol{\alpha}} \widetilde{S}_{\nu}(\mathbf{q}) 
\partial^{\boldsymbol{\beta}} \widetilde{S}_{\mu}^{\ast}(\mathbf{q}) ,
\label{eq:MagnetizationFourierCrossCorrelationFunction}
\end{align}
we rewrite the sf SANS cross section for the perpendicular scattering geometry [see Fig.~\ref{fig2} and Eq.~\eqref{eq:equation1}] as follows:
\begin{align}
    \frac{d\Sigma_{\mathrm{sf}}}{d\Omega}(\mathbf{q}) &= 
    \frac{8\pi^3}{V} b_{\mathrm{H}}^2 \left(
    \widetilde{\Gamma}^{xx}(\mathbf{q}) 
    + 
    \widetilde{\Gamma}^{yy}(\mathbf{q}) \cos^4\theta
    \right. \nonumber
    \\
    &+\left.
    \widetilde{\Gamma}^{zz}(\mathbf{q})\sin^2\theta \cos^2\theta
    -[\widetilde{\Gamma}^{yz}(\mathbf{q})+\widetilde{\Gamma}^{zy}(\mathbf{q})]
    \sin\theta \cos^3\theta
    \right),
\label{eq:SpinFlipSANS_cross_sectionWithCorrelationFunctions}
\end{align}
with $\mathbf{q} = q\{0, \sin\theta, \cos\theta \}$. In the following discussion, we focus on the second-order approximation and we neglect interparticle interaction effects.

\subsection{Second-order approximation for a dilute ensemble of spherical nanoparticles}

For a dilute ($\mathbf{a}_{\nu} = \mathbf{a}_{\mu}$) and monodisperse ($\widetilde{S}_{\nu} = \widetilde{S}_{\mu} = \widetilde{S}$) ensemble of spherical nanoparticles (with radius $R$), the Fourier cross-correlation functions simplify to~\cite{michelsbook}:
\begin{align}
\widetilde{\Gamma}^{\iota\kappa}(\mathbf{q})
&= 
i^{|\boldsymbol{\alpha}|-|\boldsymbol{\beta}|}
M_{\mu,\boldsymbol{\alpha}}^{\iota}
M_{\mu,\boldsymbol{\beta}}^{\kappa}
  \partial^{\boldsymbol{\alpha}} \widetilde{S}(q) 
\partial^{\boldsymbol{\beta}} \widetilde{S}(q),
\label{eq:FourierCrossCorrelationMatrixDilutePowerSeriesModel}
\end{align}
where $q=\sqrt{q_x^2 + q_y^2 + q_z^2}$, and
\begin{align}
 \widetilde{S}(q) = \frac{3 V_{\mathrm{s}}}{(2\pi)^{3/2}} \frac{j_1(qR)}{qR} \quad \mathrm{with}
 \quad V_{\mathrm{s}} = \frac{4\pi R^3}{3} .
\end{align}
$j_1(z)= \sin z/z^2 - \cos z/z$ is the first-order spherical Bessel function. In this special case of spherical nanoparticles (where $\widetilde{S} = \widetilde{S}^{\ast}$), the Fourier transform of the indicator function becomes real-valued, such that it is obvious that only terms with $|\boldsymbol{\alpha}|-|\boldsymbol{\beta}| = 2 u$ (with $u \in \mathbb{Z}$) contribute to Eq.~\eqref{eq:FourierCrossCorrelationMatrixDilutePowerSeriesModel}~\footnote{{This follows from the fact that the SANS cross section is a real-valued quantity.}}. In the study of Adams~et.~al.~\cite{michaeliucrj2023}, the zero-order case of Eq.~\eqref{eq:FourierCrossCorrelationMatrixDilutePowerSeriesModel}, which reflects a dilute and monodisperse ensemble of uniformly magnetized spherical nanoparticles, was studied in the context of the Stoner-Wohlfarth model. In this situation, the cross-correlation matrix can be written as: 
\begin{align}
\widetilde{\Gamma}^{\iota\kappa}(\mathbf{q})
= 
\Gamma_0^{\iota\kappa}
 [\widetilde{S}(q)]^2 
 \quad \mathrm{with}\quad 
\Gamma_0^{\iota\kappa} = 
 M_{\nu, (0,0,0) }^{\iota} M_{\nu, (0,0,0) }^{\kappa} .
\end{align}
The real-space cross-correlation matrix $\Gamma_0^{\iota\kappa}$ is a function of the applied magnetic field, such that the two-dimensional magnetic SANS cross section exhibits different types of angular anisotropies, even for randomly-averaged ensembles at remanence or at the coercive field~\cite{michaeliucrj2023}.

Now, taking into account spin inhomogeneities up to the second polynomial order in the expansion of the magnetization [Eq.~(\ref{eq:MagnetizationPowerExpansion})], Eq.~\eqref{eq:FourierCrossCorrelationMatrixDilutePowerSeriesModel} becomes:
\begin{align}
 \widetilde{\Gamma}^{\iota\kappa}(\mathbf{q})
&=  \Gamma_{0}^{\iota\kappa} \widetilde{S}^2
+
\Gamma_{1,lm}^{\iota\kappa} \left[\frac{\partial \widetilde{S}}{\partial q_{l}}\right]\left[
\frac{\partial \widetilde{S}}{\partial q_{m}}\right]
+
\Gamma_{2,lmnp}^{\iota\kappa} \left[\frac{\partial^{2} \widetilde{S}}{\partial q_{l} \partial q_{m}}\right]\left[
\frac{\partial^{2} \widetilde{S}}{\partial q_{n}\partial q_{p}}\right]
-
C_{02,lm}^{\iota\kappa}
\widetilde{S}
\left[\frac{\partial^{2} \widetilde{S}}{\partial q_{l}\partial q_{m}}\right],
\label{eq:SecondOrderApproximationMagnetizationFourierCrossCorrelationMatrix1}
\end{align}
where we have defined the combinations of polynomial magnetization coefficients as:
\begin{align}
\Gamma_0^{\iota\kappa} &= 
 M_{\nu ,\boldsymbol{\digamma}_{0}}^{\iota} M_{\nu ,\boldsymbol{\digamma}_{0}}^{\kappa},
 \label{eq:CorrCoeffDefinition1}
  \\
 \Gamma_{1,\ell m}^{\iota\kappa} &= M_{\nu,\boldsymbol{\digamma}_{\ell}}^{\iota} M_{\nu,\boldsymbol{\digamma}_{m}}^{\kappa}, \label{eq:CorrCoeffDefinition2}
  \\
 \Gamma_{2,\ell m n p}^{\iota\kappa} &= M_{\nu,(\boldsymbol{\digamma}_\ell + \boldsymbol{\digamma}_m)}^{\iota} M_{\nu,(\boldsymbol{\digamma}_n + \boldsymbol{\digamma}_p)}^{\kappa},
 \label{eq:CorrCoeffDefinition3}
 \\
 C_{02,\ell m}^{\iota \kappa} &= 
 M_{\nu,\boldsymbol{\digamma}_{0}}^{\iota}
 M_{\nu,(\boldsymbol{\digamma}_{\ell} + \boldsymbol{\digamma}_{m})}^{\kappa}
+
 M_{\nu,(\boldsymbol{\digamma}_{\ell} + \boldsymbol{\digamma}_{m})}^{\iota}
 M_{\nu,\boldsymbol{\digamma}_{0}}^{\kappa},\label{eq:CorrCoeffDefinition4}
\end{align}
and we use the following $\boldsymbol{\digamma}_{i}$~symbol for booking the multi-indices of the magnetization coefficients:
\begin{align}
\boldsymbol{\digamma}_{i} &= 
\begin{cases}
(0,0,0) &, \; i = 0
\\
(1,0,0) &, \; i = x
\\
(0,1,0) &, \; i = y
\\
(0,0,1) &, \; i = z
\end{cases}.
\label{eq:DefinitionDigammaSymbol}
\end{align}
We note that the new $\Gamma$ and $C$~coefficients include the sum over the ensemble of nanoparticles. This is seen from the fact that the index $\nu$ occurs only on the right-hand-side of Eqs.~\eqref{eq:CorrCoeffDefinition1}$-$\eqref{eq:CorrCoeffDefinition4}, but not on the left-hand-side. Since (for a spherical particle) the Fourier transform of the indicator function $\widetilde{S}$ depends only $q = \sqrt{q_x^2 + q_y^2 + q_z^2}$, we can express the partial derivatives of $\widetilde{S}$ in Eq.~\eqref{eq:SecondOrderApproximationMagnetizationFourierCrossCorrelationMatrix1}  (using the chain rule) up to the second-order as:
\begin{align}
 \widetilde{S}(q) &=  \frac{3 V_{\mathrm{s}}}{(2\pi)^{3/2}} \frac{j_1(qR)}{qR}
,&
\frac{\partial  \widetilde{S}}{\partial q_{\alpha}} &=  \hat{q}_{\alpha}   \widetilde{S}'
,&
\frac{\partial^2  \widetilde{S}}{\partial q_{\alpha} \partial q_{\beta}} &=  \hat{q}_{\alpha} \hat{q}_{\beta}   \widetilde{S}'' +
(\delta_{\alpha\beta} - \hat{q}_{\alpha} \hat{q}_{\beta}) \frac{\widetilde{S}'}{q}
,\label{eq:ChainRuleOfDerivativesOfTheSphericalShapeFunction}
\end{align}
where $\hat{q}_{l} = q_{l}/q$ (with $l = x,y,z$), $\delta^{\alpha\beta}$ is the Kronecker delta symbol, and the prime denotes the derivative with respect to the radial coordinate, i.e., $\widetilde{S}' = d\widetilde{S}/dq$ and $\widetilde{S}'' = d^2\widetilde{S}/dq^2$. Using the results from Eq.~\eqref{eq:ChainRuleOfDerivativesOfTheSphericalShapeFunction}, we can rewrite Eq.~\eqref{eq:SecondOrderApproximationMagnetizationFourierCrossCorrelationMatrix1} as follows:
\begin{align}
\widetilde{\Gamma}^{\iota\kappa}(\mathbf{q})
&= 
\Gamma_0^{\iota\kappa}
\widetilde{S}^2
+ \Gamma_{1,\ell m}^{\iota\kappa}  \hat{q}^{\ell}\hat{q}^{m}\widetilde{S}'^2 \nonumber
\\
&+
 \Gamma_{2,\ell m n p}^{\iota\kappa}
 \left[
 \hat{q}^{\ell} \hat{q}^{m}   \widetilde{S}''
 +
(\delta^{\ell m} - \hat{q}^{\ell} \hat{q}^{m}) \frac{\widetilde{S}'}{q}\right]
 \left[\hat{q}^{n} \hat{q}^{p}   \widetilde{S}''
 +
(\delta^{n p} - \hat{q}^{n} \hat{q}^{p}) \frac{\widetilde{S}'}{q}\right] \nonumber
\\
&-
C_{02,\ell m}^{\iota \kappa}  \left[\hat{q}^{\ell} \hat{q}^{m}   \widetilde{S}''
+
(\delta^{\ell m} - \hat{q}^{\ell} \hat{q}^{m}) \frac{\widetilde{S}'}{q}\right] \widetilde{S}.
\label{eq:SecondOrderApproximationMagnetizationFourierCrossCorrelationMatrix2}
\end{align}
In the above formulation, we see that the angular ($\hat{q}_{\alpha}$) dependence and the radial ($q$) dependence of the cross-correlation functions are separated in the sense of a multiplication. This is an important property that facilitates the further calculations, especially the azimuthal averaging of the magnetic SANS cross section (see below). Furthermore, inspection of Eq.~\eqref{eq:ChainRuleOfDerivativesOfTheSphericalShapeFunction} shows that the shape function $\widetilde{S}$ and its ordinary derivatives with respect to the radial coordinate $q$ also depend on the radius $R$ of the particle. Therefore, it is convenient to define the dimensionless function $f(u = qR)$ such that the shape function $\widetilde{S}$ and its derivatives can be written as follows:
\begin{align}
    f(u) &= \frac{j_1(u)}{u} = \frac{\sin u - u \cos u}{u^3},
    &
    \widetilde{S}(q) &= \frac{3 V_{\mathrm{s}}}{(2\pi)^{3/2}} f(qR),
    \\
    f'(u) &= \frac{(u^2 -3)\sin u +3 u \cos u}{u^4},
    &
    \widetilde{S}'(q) &=\frac{3 V_{\mathrm{s}} R}{(2\pi)^{3/2}} f'(qR),
    \\
    f''(u) &= \frac{(12-5u^2)\sin u + u (u^2 - 12) \cos u}{u^5},
    &
    \widetilde{S}''(q) &=\frac{3 V_{\mathrm{s}} R^2}{(2\pi)^{3/2}} f''(qR). 
\end{align}
In order to write the cross-correlation matrix [Eq.~\eqref{eq:SecondOrderApproximationMagnetizationFourierCrossCorrelationMatrix2}] in compact form, we introduce the following radial functions $g_k$ and angular functions $G_k^{\iota\kappa}$:
\begin{align}
    g_0(u) &= (f(u))^2 ,
    &
    G_0^{\iota \kappa}(\hat{\mathbf{q}}) &= h\Gamma_0^{\iota\kappa},
\label{eq:MagnetizationModel_g0}
    \\
    g_1(u) &= (f'(u))^2,
    &
    G_1^{\iota \kappa}(\hat{\mathbf{q}})&= hR^2
    \Gamma_{1,\ell m}^{\iota\kappa}  \hat{q}^{\ell}\hat{q}^{m},
    \label{eq:MagnetizationModel_g1}
    \\
    g_2(u) &= (f''(u))^2,
    &
    G_2^{\iota \kappa}(\hat{\mathbf{q}})&= hR^4
    \Gamma_{2,\ell m n p}^{\iota\kappa}\hat{q}^{\ell} \hat{q}^{m}   \hat{q}^{n} \hat{q}^{p},
    \\
    g_3(u) &= \frac{f'(u) f''(u)}{u},
    &
    G_3^{\iota \kappa}(\hat{\mathbf{q}})&=
hR^4
 \Gamma_{2,\ell m n p}^{\iota\kappa}
 \left(
 \hat{q}^{\ell} \hat{q}^{m} (\delta^{n p} - \hat{q}^{n} \hat{q}^{p})
 +
 \hat{q}^{n} \hat{q}^{p} (\delta^{\ell m} - \hat{q}^{\ell} \hat{q}^{m})
 \right),
    \\
    g_4(u) &= \frac{(f'(u))^2}{u^2},
    &
     G_4^{\iota \kappa}(\hat{\mathbf{q}})&= 
  hR^4 \Gamma_{2,\ell m n p}^{\iota\kappa}
 (\delta^{\ell m} - \hat{q}^{\ell} \hat{q}^{m})
 (\delta^{n p} - \hat{q}^{n} \hat{q}^{p}),
    \\
    g_5(u) &= f(u) f''(u),
    &
    G_5^{\iota \kappa}(\hat{\mathbf{q}})&= -hR^2
C_{02,\ell m}^{\iota \kappa}  
\hat{q}^{\ell}\hat{q}^{m},
    \\
    g_6(u) &= \frac{f(u) f'(u)}{u},
    &
     G_6^{\iota \kappa}(\hat{\mathbf{q}})&=
-hR^2
C_{02,\ell m}^{\iota \kappa}  
(\delta^{\ell m} - \hat{q}^{\ell} \hat{q}^{m}),
\label{eq:MagnetizationModel_g6}
\end{align}
where $h = \left(3V_{\mathrm{s}}/(2\pi)^{3/2}\right)^2$. This allows us to express Eq.~\eqref{eq:SecondOrderApproximationMagnetizationFourierCrossCorrelationMatrix2} as the following sum:
\begin{align}
\widetilde{\Gamma}^{\iota\kappa}(\mathbf{q})
&= 
\sum_{k=0}^{6}
G_k^{\iota \kappa}(\hat{\mathbf{q}}) \, g_k(qR).
\label{gkequation111}
\end{align}%
For completeness we provide the limit of the functions $g_i(u)$ for $u\rightarrow 0$:
\begin{align}
    \lim_{u\rightarrow 0} g_0(u) &= \frac{1}{9}
    &
    \lim_{u\rightarrow 0} g_1(u) &= 0 ,
    \\
    \lim_{u\rightarrow 0} g_i(u) &= \frac{1}{225}, \; i \in\{2, 3, 4\}
    &
    \lim_{u\rightarrow 0} g_i(u) &= -\frac{1}{45} , \; i \in \{5, 6\} .
\end{align}
The azimuthally-averaged sf SANS cross section $I_{\mathrm{sf}}(q)$ for the perpendicular scattering geometry is then obtained by a projection onto the two-dimensional detector plane, i.e., setting $\hat{\mathbf{q}} = 
    \{0,  \sin\theta ,   \cos\theta\}$ in Eq.~\eqref{gkequation111}. Substituting Eq.~\eqref{gkequation111} for the $\widetilde{\Gamma}^{\iota\kappa}(\mathbf{q})$ into Eq.~\eqref{eq:SpinFlipSANS_cross_sectionWithCorrelationFunctions} and carrying out an azimuthal average [$(2\pi)^{-1} \int_{0}^{2\pi} (...) d\theta$], we obtain [Eq.~\eqref{eq:AzimuthallyAveragedSANScrosssection_SphericalParticlesSecondOrdermaintext} in the main text]
    \begin{align}
I_{\mathrm{sf}}(q) 
= \sum_{k=0}^{6} I_{\mathrm{sf}}^k \, g_k(qR),
\label{eq:AzimuthallyAveragedSANScrosssection_SphericalParticlesSecondOrder}
\end{align}
where the $I_{\mathrm{sf}}^k$ are constant prefactors
\begin{align}
    I_{\mathrm{sf}}^k &= \frac{1}{2\pi} 
    \frac{8\pi^3 b_{\mathrm{H}}^2}{V}
     \int_{0}^{2\pi}
      \left( G_k^{xx} + G_k^{yy} \cos^4\theta  + G_k^{zz} \sin^2\theta \cos^2\theta - (G_k^{yz} + G_k^{zy}) \sin\theta \cos^3\theta \right) \; d\theta.
      \label{eq:DefinitionOfTheazimuthallyAveragedSANScrossSectionCoefficients}
\end{align}
In the perfectly saturated state, the higher-order coefficients in Eq.~\eqref{eq:AzimuthallyAveragedSANScrosssection_SphericalParticlesSecondOrder} vanish and the remaining zeroth-order term is given by:
\begin{align}
I_{\mathrm{sf}}(q; B_0\rightarrow\infty) = I_{\mathrm{sf}}^{0, \mathrm{sat}}  g_0(qR) = I_{\mathrm{sf}}^{0, \mathrm{sat}}\left[ \frac{j(qR)}{qR} \right]^2 .
\end{align}
Equation~\eqref{eq:AzimuthallyAveragedSANScrosssection_SphericalParticlesSecondOrder} is one of the central results of this paper. We note that Eq.~\eqref{eq:AzimuthallyAveragedSANScrosssection_SphericalParticlesSecondOrder} is applicable to smooth magnetization inhomogeneities and not restricted to the case of surface-anisotropy-induced spin disorder. It represents an easy-to-use fit function for azimuthally-averaged magnetic SANS cross sections (of ensembles of monodisperse and dilute spherical particles) with up to 8 free fit parameters ($R$ and the $I_{\mathrm{sf}}^k$ for $k=0...6$); $I_{\mathrm{sf}}(q)$ depends linearly on the $I_{\mathrm{sf}}^k$, but nonlinearly on the sphere radius $R$. We emphasize that although Eq.~\eqref{eq:AzimuthallyAveragedSANScrosssection_SphericalParticlesSecondOrder} has been derived for the purely magnetic sf SANS cross section, it is equally well applicable to the purely magnetic SANS cross section that might be obtained by means of unpolarized SANS measurements:~as shown e.g.\ in Refs.~\cite{bersweiler2019,michelsbook}, subtracting the nuclear and magnetic unpolarized SANS cross section at saturation from the nuclear and magnetic unpolarized SANS at a lower field (assuming a field-independent nuclear scattering) results in a purely magnetic (difference) SANS cross section that is closely related to the sf SANS (just a different combination of the magnetic Fourier components). The coefficients $I_{\mathrm{sf}}^k$ may in this case simply take on different values.

In general, the coefficients $I_{\mathrm{sf}}^k$ may depend on temperature, applied magnetic field, and on the magnetic interactions (e.g., symmetric and antisymmtric exchange, magnetic anisotropy, magnetodipolar interaction), and in particular on the radius $R$ of the nanoparticle. Therefore, in the presence of a particle-size distribution function $w(R)$, the $I_{\mathrm{sf}}^k$ become functions of $R$. One may then assume certain distribution functions for the $I_{\mathrm{sf}}^k$ (e.g., Gaussian), which would lead to an unreasonably large number of free fit parameters. Instead, a more practical approach may be to carry out a fitting procedure over $w(R)$ and to interpret the $I_{\mathrm{sf}}^k$ as ensemble-averaged quantities, which then implies that they are uniformly distributed over the particle sizes.

The zero- and first-order coefficients $I_{\mathrm{sf}}^{0}$ and $I_{\mathrm{sf}}^{1}$, as functions of the correlation coefficients $\Gamma$, are given by
\begin{align}
    I_{\mathrm{sf}}^{0} &= \frac{9 V_{\mathrm{s}}^2 b_{\mathrm{H}}^2}{8V}
    \left(8\Gamma_{0}^{xx}
    +
    3
    \Gamma_{0}^{yy}
    +    
    \Gamma_{0}^{zz}\right) ,
    \\
    I_{\mathrm{sf}}^{1} &= \frac{9 V_{\mathrm{s}}^2 R^2 b_{\mathrm{H}}^2}{16V}
    \left(
    8 \Gamma_{1,yy}^{xx}
    + 8 \Gamma_{1,zz}^{xx}
    + \Gamma_{1,yy}^{yy} 
    + 5 \Gamma_{1,zz}^{yy}
    + \Gamma_{1,yy}^{zz}
    + \Gamma_{1,zz}^{zz}
    - 2\Gamma_{1,yz}^{yz}
    - 2\Gamma_{1,zy}^{yz}
    \right) .
\end{align}
We do not provide the higher-order coefficients due to their complexity. Moreover, using the binomial inequality $a^2+b^2\ge 2ab$ ($\forall a,b\in \mathbb{R}$), it can be shown that
the coefficients $I_{\mathrm{sf}}^{0}, \; I_{\mathrm{sf}}^{1} , \; I_{\mathrm{sf}}^{2}, \; I_{\mathrm{sf}}^{4}$ are positive definite ($\in \mathbb{R}^{+}$, including zero), while the $I_{\mathrm{sf}}^{3}, \; I_{\mathrm{sf}}^{5}, \; I_{\mathrm{sf}}^{6}$ can take on positive as well as negative real values. We have additionally checked this result numerically by random sampling.

\section{Effect of core-anisotropy symmetry on the magnetic SANS cross section}
\label{appendixb}

\begin{figure}[tb!]
\centering
\resizebox{0.90\columnwidth}{!}{\includegraphics{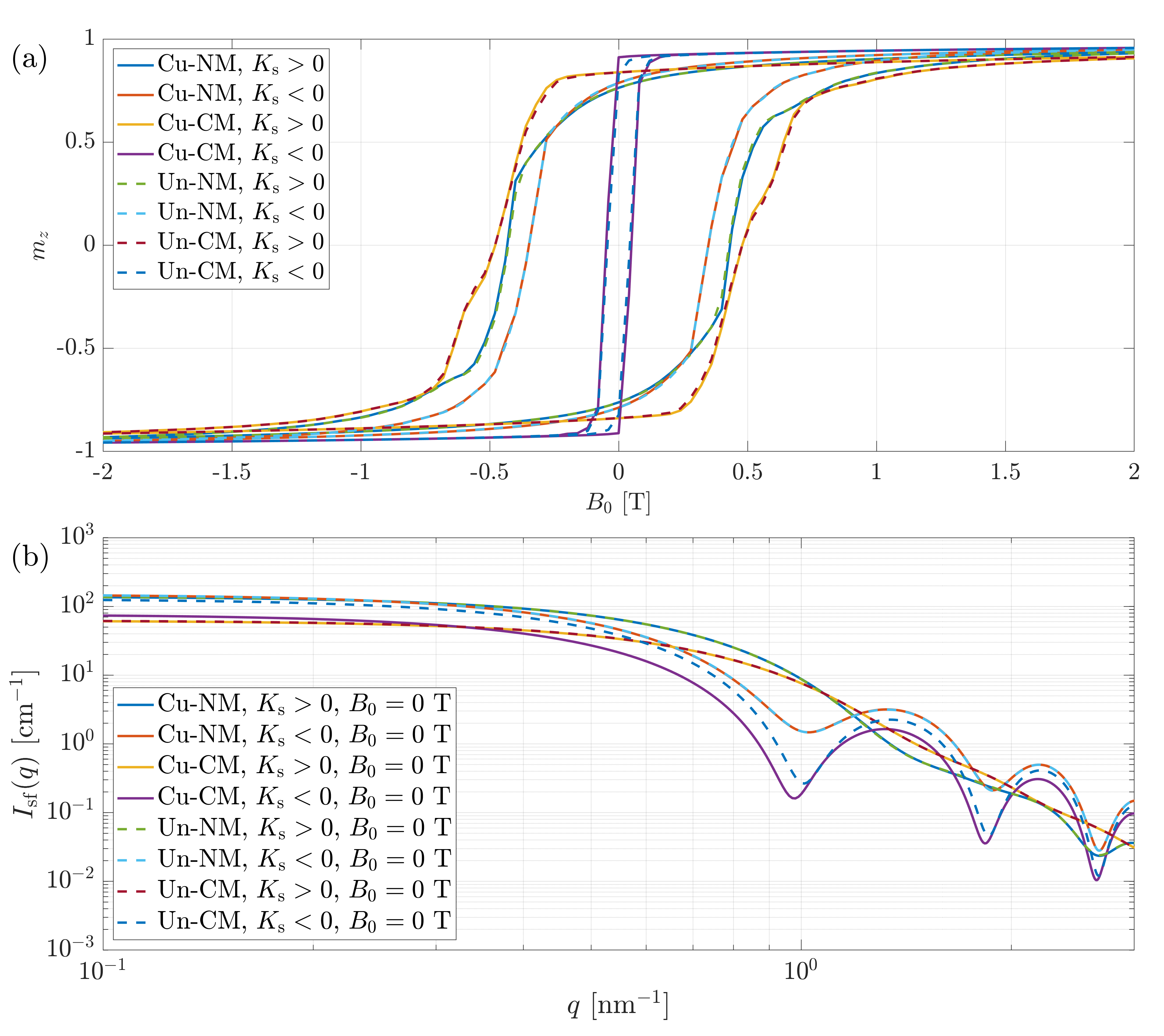}}
\caption{Effect of the symmetry of the core anisotropy (cubic ``Cu'' versus uniaxial ``Un'') on the azimuthally-averaged spin-flip SANS cross section for different combinations of the surface anisotropy constant $K_{\mathrm{s}}$. The cubic or uniaxial core anisotropy has always a magnitude of $+ 5.67 \times 10^{-25} \, \mathrm{J/atom}$, while $|K_{\mathrm{s}}| = 5.22 \times 10^{-21} \, \mathrm{J/atom}$ with the sign of $K_{\mathrm{s}}$ changing (see insets). (a)~Normalized magnetization curves $m_z(B_0)$. (b)~Azimuthally-averaged $I_{\mathrm{sf}}(q)$ at remanence (log-log scale).}
\label{fig15}
\end{figure}

Figure~\ref{fig15} displays the effect of the symmetry of the magnetic anisotropy in the core of the nanoparticles (cubic versus uniaxial) on the magnetization curve and on the azimuthally-averaged spin-flip SANS cross section $I_{\mathrm{sf}}(q)$. For a given sign of the surface anisotropy constant $K_{\mathrm{s}}$ and surface anisotropy model (N\'{e}el or conventional model), it is seen that changing the symmetry of the core anisotropy from cubic to uniaxial has only very little influence on the randomly-averaged $m_z(B_0)$ and $I_{\mathrm{sf}}(q)$. Only for the case ``Cu-CM'' and ``Un-CM'' does one see a significant difference between the curves, which might, however, be diminished in the presence of a distribution of particle sizes. We emphasize that the data in Fig.~\ref{fig15} is for a relatively large $K_{\mathrm{s}}$~value. Reducing the magnitude of the surface anisotropy will result in more homogeneous spin structures eventually approaching the Stoner-Wohlfarth results (in the limit $K_{\mathrm{s}} \rightarrow 0$)~\cite{adamsjacnum2022,michaeliucrj2023}.

\bibliography{BIB}
\bibliographystyle{apsrev4-2}

\end{document}